\documentclass[journal]{IEEEtran}

\usepackage{ucs}
\usepackage[utf8x]{inputenc}
\usepackage[cmex10]{amsmath}
\usepackage{cite, amsfonts, amssymb, amsthm, bm, bbm, graphicx, relsize, multirow, booktabs, color, blindtext}
\usepackage[american]{babel}
\usepackage[T1]{fontenc}

\theoremstyle{definition}

\usepackage{algorithmic, algorithm}
\algsetup{indent=2em, linenodelimiter=.}

\theoremstyle{remark}

\newcommand*{\herm}{^{\mathsf{H}}}
\newcommand*{\transp}{^{\mathsf{T}}}

\DeclareMathOperator{\card}{card}
\DeclareMathOperator{\diag}{diag}
\DeclareMathOperator{\trace}{Tr}
\DeclareMathOperator*{\argmax}{\arg\,\max}

\newcommand{\e}{\mathrm{e}}
\renewcommand{\i}{\mathrm{i}}

\IEEEoverridecommandlockouts

\title{Energy Efficiency Optimization in Radar-Communication Spectrum Sharing}

\author{Emanuele~Grossi,~\IEEEmembership{Senior Member,~IEEE,}
 Marco~Lops,~\IEEEmembership{Fellow,~IEEE,}
 and Luca~Venturino,~\IEEEmembership{Senior Member,~IEEE}%
\thanks{E. Grossi and L. Venturino are with the Department
of Electrical and Information Engineering (DIEI), Universit\`a degli Studi di Cassino e del Lazio Meridionale, Italy 03043, and with Consorzio Nazionale Interuniversitario per le Telecomunicazioni (CNIT), Parma, Italy 43124; e-mail: e.grossi@unicas.it, l.venturino@unicas.it.}%
\thanks{M. Lops is with the Department of Electrical Engineering and Information Technology (DIETI), Universit\`a degli Studi di Napoli ``Federico II,'' Italy 80125, and with Consorzio Nazionale Interuniversitario per le Telecomunicazioni (CNIT), Parma, Italy 43124; email: lops@unina.it.}%
}

\begin{document}
\bstctlcite{BSTcontrol}
\maketitle

\begin{abstract}
Energy efficiency, possibly coupled with cognition-based and spectrum-sharing architectures, is a key enabling technology for green communications in 5G-and-beyond standards. In this context, the present paper considers a multiple-input multiple-output communication system cooperatively coexisting with a surveillance radar: the objective function is the communication system energy efficiency, while radar operation is safeguarded by constraining the minimum received signal-to-disturbance ratio for a set of range-azimuth cells of the controlled scene, and no time synchronization between them is assumed. The degrees of freedom are the transmit powers of both systems, the space-time communication codebook and the linear filters at the radar receiver. The resulting optimization problem is non-convex, due to both the objective function and the presence of signal-dependent interference (clutter): we develop a block-coordinate-ascent approximate solution, and offer a thorough performance assessment, so as to elicit  the merits of the proposed approach, along with the interplay among the achievable energy efficiency, the density of scatterers in the environment, and the size of the set of protected radar cells.
\end{abstract}


\section{Introduction}
The exponential increase of data traffic in wireless networks in recent years and the corresponding growth in infrastructure has led to a rapid increase of the energy consumed by wireless communication systems~\cite{Lambert_2012}. The massive data rates required by a pervasive connectivity cannot be guaranteed by simply scaling up the transmit power, since it would inevitably result in an unmanageable energy demand; on the contrary, the energy consumption should be reduced~\cite{5g-ppp}. The Energy Efficiency (EE) of the communication systems should therefore be increased, and, indeed, this new metric has recently been included in the International Telecommunication Unit recommendations for IMT-2020 radio interfaces~\cite{ITU_2017}. A basic definition of EE, measured in Joule/second, is the ratio of the data rate (in bit/second) to the power consumption (in Watt)~\cite{Kwon_1986, Isheden_2012}, where the denominator includes the transmit power, on top of the power dissipated in the transceiver hardware and baseband processing~\cite{Mammela_2017}.

In recent years, various strategies have been proposed to reduce the power consumption of both mobile and fixed networks at all levels, from the access network to the core~\cite{Correia_2010, Chen_2011, Vereecken_2011, Hinton_2011, Han_2011, Bianzino_2012, Feng_2013}:  this is of paramount importance for operators (e.g., to reduce the operating expenditure costs) and for end-users (e.g., to prolong the battery lifetime). The optimization of the EE has been the focus of an intense research in uplink~\cite{Miao_2012} and~downlink~\cite{Xiong_2011, Ng_2013} Orthogonal Frequency Division Multiple access (OFDMA) systems, Non-Orthogonal Multiple Access (NOMA) networks~\cite{Fang_2016}, heterogeneous cellular networks~\cite{Soh_2013}, Multiple-Input Multiple-Output (MIMO)~\cite{Xu_2013, Nguyen_2013} and massive MIMO~\cite{Ngo_2013, Bjornson_2014} systems---possibly exploiting Reconfigurable Intelligent Surfaces (RIS)~\cite{Huang_2019}---, cognitive radio networks~\cite{Ren_2016}, Device-to-Device (D2D) communications~\cite{Fodor_2012, Feng_2014}, Internet of Things (IoT)~\cite{Wang_2016, Zhang_2016}, and 5G networks~\cite{Buzzi_2016, Zappone_2016}.

The optimization of the EE thus appears to be a very promising answer to the explosive growth of connected devices and the tremendous increase in the data traffic~\cite{QUALCOMM_2013, CISCO_internet_2020}, but the frequency spectrum is nevertheless becoming increasingly congested. The network providers are meanwhile seeking opportunities to reuse frequencies currently restricted to other applications, and the radar bands appear to be one of the best choices, given the large portions of spectrum available to these systems~\cite{Griffiths_2015}. It is in fact anticipated that, in the near future, the 2--8~GHz frequency range, comprising traditional S and C radar bands, will inevitably experience a cohabitation with wireless communication systems, e.g., 5G New Radio (NR), Long Term Evolution (LTE), and Wi-Fi~\cite{Griffiths_2015, Liu_2020}. Higher frequencies, such as the mmWave band, used, e.g., by automotive radars and for high-resolution imaging, are also supposed to be shared for communication and sensing and achieve harmonious coexistence or even beneficial cooperation in the 5G network and beyond. Accordingly, the Defense Advanced Research Projects Agency (DARPA) recently announced the shared spectrum access for radar and communications (SSPARC) program~\cite{SSPARC_2013, Jacyna_2016}.

A number of approaches aimed at assessing the feasibility of spectrum sharing have been proposed so far; a possible classification might follow the taxonomy proposed in~\cite{Liu_2020}, wherein the major categorization is between Radar-Communication Coexistence (RCC) and Dual-Functional Radar-Communication (DFRC) architectures. In the latter case, the two systems are combined in the same hardware platform, which is designed to guarantee the performance of both systems by fully exploiting the higher degree of information sharing granted by colocation~\cite{Chiryath_2016, Liu_2018}. Starting from the pioneering study in~\cite{Mealey_1963}, that put forward the idea of modulating communication bits onto radar pulses, a number of works have been proposed for embedding communication information into radar signal both in time~\cite{Blunt_2010} and in space~\cite{Hassanien_2016, Wang_Hassanien_2019}, and for using communication signals to perform sensing operations~\cite{Sturm_2011, Zhou_2019}, possibly at mmWaves~\cite{Grossi_2017_radarcon, Grossi_2017_VTC, Grossi_2018_TSP, Kumari_2018, Grossi_2019_asilomar, Grossi_2019_camsap_b, Grossi_2020_TWC}. In the RCC case, instead, interference management techniques and joint/disjoint transceiver design have been advocated so that the two systems can safely operate in the same frequency band. Many approaches have been proposed so far entailing different degrees of cooperation. In~\cite{Wang_2008, Saruthirathanaworakun_2012, Hessar_2016, Raymond_2016} strict policies for coordinated spectrum access are defined, possibly using channel sensing techniques borrowed from the large cognitive radio literature, while interference channel estimation is exploited in~\cite{Li_2017, Liu_2019}. The studies in~\cite{Deng_2013, Geng_2015, Mahal_2017, Cheng_2018, Shi_2018, Bica_2019} focus on the radar system, whose transmitted waveform is carefully designed, while in~\cite{Nartasilpa_2018} the communication system is the primary element of concern, and its receiver is accurately designed. Increased degrees of freedom/cooperation obviously allow achieving better results, and in~\cite{Li_2016, Zheng_2017, Zheng_2018_a, Grossi_2018, Rihan_2018, Grossi_2019_camsap_a, Wang_2019, Cheng_2019, Grossi_2020_TSP} optimization-based joint-design (or co-design) of the radar waveform and of the communication system codebook is undertaken.

In this work, starting from the preliminary results presented in~\cite{Grossi_2020_SAM}, we optimize the EE of a MIMO communication system in the presence of the interference caused by a surveillance radar. The contest is that of RCC, and leverages the signal model presented in~\cite{Grossi_2020_TSP} to undertake joint system design. The performance of the radar is guaranteed by forcing the received Signal-to-Disturbance Ratio (SDR) at each resolution bin to exceed a prescribed level: this is a key requirement for surveillance radars, where the monitored area is wide, and all the observed resolution cells should be protected by excessive interference. The degrees of freedom for system optimization are the transmit power and the receive filters of the radar, while, for the communication system, it is the whole codebook, represented by the covariance matrix of the Space-Time Codewords (STC's).

The major novelty of the present study is the optimization of the EE in the presence of the interference caused by a radar system: at the best of authors' knowledge, this problem has not been considered so far. The resulting optimization problem is quite complex, whereby we propose a block coordinate ascent method (also known as alternating maximization) combined with Dinkelbach's algoritm and with a quasi-Newton projected gradient method to find an approximate solution thereof. The analysis demonstrates that large gains with respect to a disjoint design of the two systems can be achieved in terms of both EE of the communication system and SDR level at the radar side, and that there is a key tradeoff among EE, density of clutter scatterers in the surrounding environment, and number of protected radar cells.

The rest of the paper is organized as follows. In the next section, the signal model at the radar and communication system sides is introduced. In Sec.~\ref{joint_opt_sec} the joint optimization problem is presented and (sub-optimally) solved, with mathematical derivations deferred in the Appendix, while in Sec.~\ref{num_ex_sec} a numerical example is provided to show the merits of the proposed strategy. Finally, some concluding remarks are given in Sec.~\ref{conclusion}.

\paragraph*{Notation} In the following, $\mathbb R$, $\mathbb R_+$, and $\mathbb C$ denote the set of real, non-negative real, and complex numbers, respectively; $\i$ is the imaginary unit, while $\Re\{\,\cdot\,\}$ and $\Im\{\,\cdot\,\}$ are the real and imaginary parts, respectively; $\bm X\succ 0$ and $\bm X\succeq0$, means that $\bm X$ is Hermitian positive definite and positive semidefinite, respectively; $x^+=\max\{x,0\}$ is the positive part of $x\in\mathbb R$, while $\bm X^+$ is the projection of the matrix $\bm X$ onto the cone of Hermitian positive semidefinite matrices; $(\,\cdot\,)^*$, $(\,\cdot\,)\transp$, and $(\,\cdot\,)\herm$ denote conjugate, transpose, and conjugate transpose; $\bm I_n$ is the $n\times n$ identity matrix; $\bm 0_n$ is the all-zero $n$-dimensional column vector; $\bm O_{m,n}$ is the $m\times n$ matrix with all zero entries; $\diag(a_1,\ldots,a_n)$ is the $n\times n$ diagonal matrix with entries $\{a_i\}_{i=1}^n$ on the principal diagonal; $\trace \bm X $ is the trace of the matrix $\bm X$; $\Vert \,\cdot\,\Vert$ denotes the Euclidean norm; $[\bm X]_{i:j,h:k}$ is the sub-matrix consisting of the rows $i$ through $j$ and the columns $h$ through $k$ of the matrix $\bm X$; $\bm X^{1/2}$ is the square root matrix of $\bm X \succeq0$, while $\bm X^{-1/2}$ is the square root matrix of the inverse of $\bm X \succ0$; $\mathbb E [\, \cdot\,]$ denotes statistical expectation; $\mathcal N_c(\bm 0, \bm X)$ is the complex circularly-symmetric Gaussian distribution with covariance matrix $\bm X$; and $\card \mathcal S$ is the cardinality of the set $\mathcal S$.

\section{Signal model}

We consider a MIMO communication system coexisting with a surveillance radar on the same bandwidth $W$. The communication system is embedded in a local rich scattering environment (e.g., an urban area) with approximate size $c/W$, $c$ denoting the speed of light. The radar monitors a large region, that includes the one where the communication system operates; since the bandwidth is $W$, the radar range resolution is in the order of $c/(2W)$~\cite{Skolnik_2001}. Both systems adopt a linear modulation.

The radar illuminates all the inspected area with a non-scanning wide-beam transmit antenna.\footnote{Non-scanning radars employ a broad-beam antenna covering the whole search area and are in general complemented at the receiver by a proper azimuth sectorization through multiple narrow beams. They are also known as floodlight radars, since the search area is \emph{flooded} with the transmitted signal~\cite{Rudge_1983, Wirth_2013}, or \emph{ubiquitous} radars, since the search area is constantly monitored, and different tasks can simultaneously be performed at the receiver~\cite{Skolnik_2002_a}.} It emits an encoded pulse train with average transmit power $P_r$ and pulse repetition time (PRT) $T$, so that the number of \emph{non-ambiguous} range cells is $N\approx WT$~\cite{Skolnik_2001}. The (fast-time) code sequence, used to modulate the subpulses composing each pulse is $\bm q=[q(0)\, \cdots \,q(L-1)]\transp\in \mathbb C^L$, where $0<L<N$, and $\bm q$ is normalized so that $\Vert \bm q \Vert^2=N$. The communication transmitter is equipped with $M$ omni-directional antennas, and the symbols emitted by all the antennas during $N$ signaling intervals form an $M\times N$ STC: the $p$-th STC is, therefore, $\{c_m(pN+i): m=1,\ldots, M, i=0,\ldots, N-1\}$, where $c_m(i)$ is the symbol sent at epoch $i$ from antenna $m$. We assume that STC's emitted at different epochs are independent.

The communication receiver is equipped with $K$ omni-directional antennas, and the discretized version of the received communication signal corresponding to the $p$-th codeword can be written as~\cite[Sec.~II-A]{Grossi_2020_TSP}
\begin{equation}
\bm r = \underbrace{(\bm H \otimes \bm I_N) \bm c}_{\text{signal of interest}}+ \underbrace{\sqrt{P_r} \sum_{i=0}^{N-1} \bm \alpha_i \otimes \bm q_i}_{\text{radar inteference}} +\underbrace{\bm v \label{r_comm_discrete}}_{\substack{\text{thermal}\\ \text{noise}}} \in \mathbb C^{KN}
\end{equation}
where:
\begin{itemize}
 \item $\bm H\in\mathbb C^{K\times M}$ is the channel matrix, whose $(k,m)$ entry is the gain of the channel linking receive antenna $k$ to transmit antenna $m$; $\bm H$ is assumed to be perfectly estimated;
 \item $\bm c=[\bm c_1\transp \, \cdots \, \bm c_M\transp]\transp \in \mathbb C^{MN}$ is the $p$-th STC, where $\bm c_m= [c_m(0) \, \cdots \, c_m(N-1)]\transp \in \mathbb C^N$ is the codeword emitted by antenna $m$;
 \item $\bm \alpha_i\sim \mathcal N_c(\bm 0_{K}, \bm \Sigma_{\alpha,i})$ is the vector containing the amplitudes of the radar echoes hitting the receive antennas during the $i$-th signaling interval;\footnote{Notice that some of the covariance matrices $\{\bm \Sigma_{\alpha,i}\}_{i=1}^N$ may be equal to $\bm O_{K,K}$, two relevant special  cases being that only one of them is non-zero (e.g., the direct path) and that all of them are zero (corresponding to \emph{isolated}, i.e., non interfering systems).} we assume that $\bm \Sigma_{\alpha,i}$ has been perfectly estimated, and that $\bm \alpha_i$ and $\bm \alpha_j$ are independent for $i\neq j$;
 \item $\bm q_i\in\mathbb C^N$ is the vector obtained by taking a downwards circular shift of $i$ positions of $[\bm q\transp \, 0 \, \cdots \, 0]\transp \in \mathbb C^N$; and
 \item $\bm v \sim \mathcal N_c(\bm 0_{KN}, P_v \bm I_{KN})$ is the noise vector, with $P_v$ the (perfectly estimated) noise power.
\end{itemize}

The radar receiver forms $J$ simultaneous orthogonal beams\footnote{There are many alternative ways to form simultaneous orthogonal beams: e.g., $J$ narrow-beam antennas, $J^2$ analog phase shifters in an antenna array with $J$ (or more) radiating elements, digital beamforming through $J$ A/D converters in the same antenna array, etc. These beams will not be perfectly orthogonal, due to the inevitable presence of sidelobes, but the attenuation can be made quite large, and we can assume, at the design stage, that the signal term from the sidelobes can be neglected.} so as to cover the area illuminated by the non-scanning wide transmit beam; the number of azimuth bins is, therefore, $J$. Assuming the presence of a point-like target with delay\footnote{Delays smaller than the duration of the emitted pulse correspond to distances of no interest to the radar.} $(L+n)T_c$, $n\in\{0,\dots,N-L\}$, in the $j$-th azimuth bin, the discretized version of the received radar signal from the $j$-th beam can be written as~\cite[Sec.~II-B]{Grossi_2020_TSP}
\begin{align}
 \bm y_j&= \underbrace{\sqrt{P_r} g_{n,j} \bm q_n}_{\text{target echo}} + \underbrace{\sqrt{P_r} \sum_{i=0}^{N-1} \gamma_{i,j} \bm q_i}_{\text{radar clutter}} \notag\\
 & \quad + \underbrace{\sum_{m=1}^M \sum_{i=0}^{N-1} \sum_{d=0}^\infty \beta_{m,i,j,d} \bm c_{m,i,d}}_{\text{data inteference}} + \underbrace{\bm u_j}_{\substack{\text{thermal}\\ \text{noise}}} \label{r_radar_discrete}
\end{align}
where:
\begin{itemize}
 \item $g_{n,j}\in \mathbb C$ is the amplitude of the target echo, modeled as a zero-mean random variable with variance $\sigma^2_{g,n,j}$;
 \item $\gamma_{i,j}\in \mathbb C$ is the amplitude of the clutter echoes in the range-azimuth bin $(i,j)$, modeled as a zero-mean random variable with variance $\sigma^2_{\gamma,i,j}$; we assume that $\sigma^2_{\gamma,i,j}$ has been perfectly estimated, and that $\gamma_{i,j}$ and $\gamma_{i',j'}$ are independent for $(i,j)\neq (i',j')$;
 \item $\beta_{m,i,j,d}\in \mathbb C$ is the amplitude of the communication echoes from antenna $m$ pertaining to the range-azimuth bin $(i,j)$; the subscript $d$ denotes echoes separated by a delay $dT$ that, due to the radar range ambiguity, correspond to the same range bin; we model it as a zero-mean random variable,\footnote{In practice, only the signals reflected by objects located at a sufficiently short range and having a sufficiently large radar cross-section are received with a non-negligible power, so that only few of the coefficients $\{\beta_{m,i,j,d}\}_{d=0}^\infty$ (mostly, only the term $d=0$) have a non-zero variance. Furthermore, similarly to the communication receiver case, some of the terms $\{\beta_{m,i,j,d}\}_{i=0}^{N-1}$ may have a zero variance, the limiting cases being those where only one has a non-zero variance (e.g., the direct path) and where all of them have variance zero (corresponding to \emph{isolated}, i.e., non interfering systems).} and we assume that $\beta_{m,i,j,d}$ and $\beta_{m',i',j',d'}$ are independent for $(i,j,d)\neq(i',j',d')$; also, we assume that $\sigma^2_{\beta,m,m',i,j}= \sum_{d=0}^\infty \mathbb E[ \beta_{m,i,j,d}\beta_{m',i,j,d}^*]$ has been perfectly estimated;
 \item $\bm c_{m,i,d}$ is the sequence of $N$ symbols transmitted by antenna $m$ of the communication system that fall in the $p$-th PRT, i.e.,
\begin{equation}
\bm c_{m,i,d}=\begin{bmatrix} c_m\big((p-d)N- \nu_0+L-i\big) \\ \vdots \\ c_m\big((p-d)N-\nu_0+L-i +N-1\big)\end{bmatrix}
\end{equation}
with $\nu_0$ denoting the delay of the communication transmitter with respect to the radar transmitter (i.e., the asynchrony between the two systems),\footnote{If $\tau$ is the smallest travelling time between the communication transmitter and the radar receiver, and $\tau'$ is the delay (if any) of the communication transmitter with respect to the radar transmitter, then $\nu_0$ is equal to $\tau+\tau'$ expressed in symbol intervals.} assumed to be perfectly estimated; and
\item $\bm u_j \sim \mathcal N_c(\bm 0_N, P_u \bm I_N)$ is the noise vector, with $P_u$ the (perfectly estimated) noise power.
\end{itemize}
Notice that $\bm c_{m,i,d}$ contains the last $\ell_i=(\nu_0-L+i) \mod N$ symbols of the $m$-th segment of the codeword $p-d- \big \lfloor (\nu_0-L+i)/N \big \rfloor-1$ and the first $N-\ell_i$ symbols of the $m$-th segment of the codeword $p-d- \big \lfloor (\nu_0-L+i)/N \big \rfloor$. Hence, the cross-covariance matrix of $\bm c_{m,i,d}$ and $\bm c_{m',i,d}$ is
\begin{align}
&\bm C_{m,m',i}=\mathbb E\left[ \bm c_{m,i,d} \bm c_{m',i,d}\herm\right] \notag\\
&= \begin{bmatrix}
[\bm C_{m,m'}]_{N-\ell_i+1:N,N-\ell_i+1:N} & & \bm O_{\ell_i,N-\ell_i} 
\\[10pt]
\bm O_{N-\ell_i,\ell_i} & & [\bm C_{m,m'}]_{1:N-\ell_i,1:N-\ell_i}
\end{bmatrix} \notag\\
&=\bm A_{m,i} \bm C \bm A_{m',i}\transp + \bm B_{m,i} \bm C \bm B_{m',i}\transp\label{partial_cov_expr}
\end{align}
where $\bm C_{m,m'}=\mathbb E\left[ \bm c_m \bm c_{m'}\herm\right]$, $\bm C=\mathbb E\left[ \bm c \bm c\herm\right]$, and
\begin{subequations}
\begin{align}
 \bm A_{m,i}& = \begin{bmatrix}
 \bm O_{\ell_i,(m-1)N} & \bm O_{\ell_i,N-\ell_i} & \bm O_{\ell_i,(M-m)N+\ell_i} \\
 \bm O_{N-\ell_i, (m-1)N} & \bm I_{N-\ell_i} & \bm O_{N-\ell_i,(M-m)N+\ell_i}
 \end{bmatrix}\\
 \bm B_{m,i}& = \begin{bmatrix}
 \bm O_{\ell_i,mN-\ell_i} & \bm I_{\ell_i} & \bm O_{\ell_i,(M-m)N} \\
 \bm O_{N-\ell_i, mN-\ell_i} & \bm O_{N-\ell_i,\ell_i} & \bm O_{N-\ell_i,(M-m)N}
 \end{bmatrix}.
\end{align}%
\end{subequations}

\section{Joint optimization}\label{joint_opt_sec}

Here we tackle the problem of joint transceiver design. As to the degrees of freedom for system optimization, we assume an optimum (capacity-achieving) architecture at the receive side of the communication system. At the transmitter, the STC's should be drawn from a complex circularly symmetric Gaussian distribution, so as to maximize the EE, since the channel in~\eqref{r_comm_discrete} is Gaussian; therefore, their covariance matrix $\bm C$ must be optimized, the transmit power being determined by its trace. At the radar side, the transmit waveform must comply with a number of requirements concerning resolution, variations in the signal modulus, sidelobe level, and ambiguity~\cite{Skolnik_2008, Deng_2004}, whereby few degrees of freedom remain once all of these constraints are taken into account. We therefore assume that the radar waveform is fixed, and we only allow the power $P_r$ to be adjusted so as to meet the specific requirements in terms of detection and estimation accuracy. At the receiver, the radar filters can be freely optimized, for they simply entail low-complexity digital signal processing operations.

The figures of merit used for system optimization are the EE of the communication system and the SDR in each resolution cell of the radar. In particular, EE is defined as the ratio of the achievable rate, $R$ (in bits per second), to the power consumption (measured in Watts), i.e.,
\begin{equation}
 \text{EE} = \frac{R}{P_c/\eta+\omega} \label{EE_expr}
\end{equation}
and is measured in bits per Joule; in the previous equation, $\eta$ is the power amplifier efficiency and $\omega$ is the circuit power required to operate the link~\cite{Kwon_1986, Isheden_2012, Bjornson_2014, Zappone_2015}. As to $R$, it is equal to the mutual information (in bits per channel use) of the Gaussian channel described in~\eqref{r_comm_discrete} times the channel bandwidth, i.e.,
\begin{align}
 R &= \frac{W}{N}\log_2 \det \left(\vphantom{\left(P_r \sum_{i=0}^{N-1} \bm \Sigma_{\alpha,i} \otimes (\bm q_i\bm q_i\herm) +P_v \bm I_{NK} \right)^{-1}} \bm I_{NK} + (\bm H \otimes \bm I_N) \bm C (\bm H \otimes \bm I_N)\herm \right. \notag\\
 &\quad \left. \times \left(P_r \sum_{i=0}^{N-1} \bm \Sigma_{\alpha,i} \otimes (\bm q_i\bm q_i\herm) +P_v \bm I_{NK} \right)^{-1}\right). \label{R_expr}
\end{align}
This is an upper bound to the achievable transmission rate, that can be approached provided $N$ is large enough. At the communication side, the covariance matrix $\bm C$ should be chosen so as to maximize the EE.

At the radar side, if pulse compression at range bin $n$ and azimuth bin $j$ is performed with the linear filter $\bm w_{n,j}$, the corresponding output is
\begin{align}
 \bm w_{n,j}\herm\bm y_j &=\sqrt{P_r} g_{n,j} \bm w_{n,j}\herm \bm q_n + \sqrt{P_r } \sum_{i=0}^{N-1} \gamma_{i,j} \bm w_{n,j}\herm \bm q_i \notag\\
 &\quad + \sum_{m=1}^M \sum_{i=0}^{N-1} \sum_{d=0}^\infty \beta_{m,i,j,d} \bm w_{n,j}\herm \bm c_{m,i,d} + \bm w_{n,j}\herm \bm u
\end{align}
and the SDR at resolution cell $(n,j)$ is, from~\eqref{r_radar_discrete} and~\eqref{partial_cov_expr},
\begin{align}
 \text{SDR}_{n,j}& =P_r \sigma^2_{g,n,j} |\bm w_{n,j}\herm\bm q_n|^2 \left[\bm w_{n,j}\herm \left( P_r\sum_{i=0}^{N-1} \sigma^2_{\gamma,i,j} \bm q_i \bm q_i\herm \right. \right.\notag\\
 & \quad + \sum_{i=0}^{N-1} \sum_{m=1}^M \sum_{m'=1}^M \sigma^2_{\beta,m,m',i,j} (\bm A_{m,i} \bm C \bm A_{m',i}\transp \notag\\
 & \quad \left. \left. \vphantom{\sum_{i=0}^{N-1}} + \bm B_{m,i} \bm C \bm B_{m',i}\transp) +P_u\bm I_N \right)\bm w_{n,j}\right]^{-1}. \label{SDR_def}
\end{align}
The radar is normally designed to guarantee a minimum level of SDR at each resolution bin, so as to be able to detect targets with specified radar cross-sections at specified locations (i.e., with specified $\sigma^2_{g,n,j}$); we denote by $\rho_{n,j}$ the minimum required SDR at resolution cell $(n,j)\in \mathcal X$, where $\mathcal X\subseteq \{0,\ldots, N-L\} \times \{1,\ldots,J\}$ is the set of radar resolution cells under observation.

The optimization problem to be solved is, therefore,
\begin{equation}
 \begin{aligned}
 \max_{\substack{\bm C \in \mathbb C^{MN\times MN} \\ \{ \bm w_{n,j}\}_{(n,j)\in\mathcal X} \in \mathbb C^N \\ P_r\in\mathbb R}} & \quad \text{EE} ( \bm C, P_r )\\
 \text{s.t.} & \quad \bm C \succeq 0\\
 & \quad \frac1N \trace \bm C \leq P_{c,\text{max}}\\
 & \quad \text{SDR}_{n,j} (\bm C, P_r, \bm w_{n,j}) \geq \rho_{n,j}, \forall (n,j) \in \mathcal X\\
 & \quad 0\leq P_r \leq P_{r,\text{max}}
 \end{aligned} \label{opt_prob}
\end{equation}
where $P_{r,\text{max}}$ and $P_{c,\text{max}}$ are the maximum transmit powers at the radar and at the communication system, respectively. We also assume that
\begin{multline}
 \sigma^2_{g,n,j} P_{r,\text{max}} \bm q_n\herm \left(P_{r,\text{max}} \sum_{i=0}^{N-1} \sigma^2_{\gamma,i,j} \bm q_i \bm q_i\herm + P_u\bm I_N \right)^{-1} \\
 \times \bm q_n > \rho_{n,j}, \quad \forall (n,j)\in\mathcal X \label{rho_max}
\end{multline}
which amounts to requiring that all the SDR constraints be satisfied at least when the communication system is transmitting with an arbitrarily small power, the radar uses all the available power, and the radar receive filters are chosen so as to maximize the SDR at each resolution cell.

Problem~\eqref{opt_prob} is quite complex, so that we resort to the block coordinate ascent method~\cite{Bertsekas_1999}, also known as non-linear Gauss-Seidel method or as alternating maximization: starting from a feasible point, the objective function is maximized with respect to each of the ``block coordinate'' variables, taken in cyclic order, while keeping the other ones fixed at their previous values.\footnote{If all maximizations are optimally solved (or, at least, the objective function is non-decreasing in successive maximizations), the algorithm converges. However, since Problem~\eqref{opt_prob} is not convex, and the feasible set cannot be expressed as the Cartesian product of closed convex sets, there is no guarantee that a global maximum is reached.} Here, the natural block coordinate variables are the radar receive filters $\{ \bm w_{n,j}\}_{(n,j)\in\mathcal X}$, the radar transmit power $P_r$, and the communication codebook $\bm C$. We solve these reduced complexity sub-problems in Secs.~\ref{radar_sol} and~\ref{comm_syst_sol}, while, in Sec.~\ref{algorithm_sec}, we report the complete algorithm, along with a discussion on its computational complexity.

\subsection{Radar optimization}\label{radar_sol}

Since the EE in~\eqref{EE_expr} is independent of the radar filters, we select $\{\bm w_{n,j}\}_{(n,j)\in\mathcal X}$ so as to maximize the SDR in each resolution cell, and, therefore, guarantee the largest feasible set. Following~\cite[Sec.~III-A]{Grossi_2020_TSP}, we have
\begin{align}
 \bm w_{n,j}&\propto \left(P_r\sum_{i=0}^{N-1} \sigma^2_{\gamma,i,j} \bm q_i \bm q_i\herm + \sum_{i=0}^{N-1} \sum_{m=1}^M \sum_{m'=1}^M \sigma^2_{\beta,m,m',i,j} \right. \notag\\
 & \quad \left.\vphantom{\sum_{i=0}^{N-1}} \times \big( \bm A_{m,i} \bm C \bm A_{m',i}\transp+ \bm B_{m,i} \bm C \bm B_{m',i}\transp \big) +P_u\bm I_N\right)^{-1} \bm q_n \label{w_update}
 \end{align}
for $(n,j)\in\mathcal X$. Furthermore, since, the EE is strictly decreasing with the radar transmit power, we must select the smallest value of $P_r$ satisfying all the SDR constraints. From~\cite[Sec.~III-B]{Grossi_2020_TSP}, we have
\begin{align}
 P_r& = \max_{(n,j)\in\mathcal X} \rho_{n,j} \left( \sum_{i=0}^{N-1} \sum_{m=1}^M\sum_{m'=1}^M \sigma^2_{\beta,m,m',i,j} \bm w_{n,j}\herm \right.\notag\\
 & \quad \times \big( \bm A_{m,i} \bm C \bm A_{m',i}\transp+ \bm B_{m,i} \bm C \bm B_{m',i}\transp \big) \bm w_{n,j} +P_u \Vert \bm w_{n,j}\Vert^2 \Bigg)\notag\\
 &\quad \times \left( \sigma^2_{g,n,j} |\bm w_{n,j}\herm \bm q_n|^2 - \rho_{n,j} \sum_{i=0}^{N-1} \sigma^2_{\gamma,i,j} |\bm w_{n,j}\herm \bm q_i|^2\right)^{-1}.\label{Pr_update}
\end{align}

Notice that, in~\eqref{w_update}, we have the inverse of the covariance matrix of the disturbance (noise plus clutter and data interference); the receiver can therefore be interpreted as the cascade of a whitening filter followed by a filter matched to the useful signal. The power level in~\eqref{Pr_update}, instead, can simply be obtained by elaborating the SDRs in~\eqref{SDR_def}, where, at the numerator, we have the useful signal (that is a linear function of $P_r$) and, at the denominator, the noise, clutter (that is a linear function of $P_r$), and data interference terms.

\subsection{Communication system optimization} \label{comm_syst_sol}

Let $\phi:\mathcal X \rightarrow \big\{1,\ldots,\card \mathcal X \big\}$ be a one-to-one mapping, $U=\card \mathcal X +1$, and
\begin{subequations}
\begin{align}
 \bm F &= (\bm H \otimes \bm I_N)\herm \left(P_r \sum_{i=0}^{N-1} \bm \Sigma_{\alpha,i}\otimes( \bm q_i\bm q_i\herm) +P_v \bm I_{KN} \right)^{-1} \notag\\
 &\quad \times (\bm H \otimes \bm I_N) \in \mathbb C^{MN\times MN} \label{mat_F}\\
\bm E_{\phi(n,j)} & = \sum_{i=0}^{N-1} \sum_{m=1}^M \sum_{m'=1}^M \sigma^2_{\beta,m,m',i,j} (\bm A_{m',i}\transp \bm w_{n,j} \bm w_{n,j}\herm \bm A_{m,i} \notag \\
& \quad + \bm B_{m',i}\transp \bm w_{n,j} \bm w_{n,j}\herm\bm B_{m,i} )\in \mathbb C^{MN\times MN} , \; (n,j)\in\mathcal X\\
 \bm E_U &=\bm I_{MN}\\
 a_{\phi(n,j)} &= \frac{P_r \sigma^2_{g,n,j}}{\rho_{n,j}} |\bm w_{n,j}\herm\bm q_n|^2 - P_r \sum_{i=0}^{N-1} \sigma^2_{\gamma,i,j} |\bm w_{n,j}\herm \bm q_i|^2\notag \\
&\quad -P_u\Vert \bm w_{n,j}\Vert^2 \geq0, \quad (n,j)\in\mathcal X\\
a_U &= N P_{c,\text{max}}.
\end{align}%
\end{subequations}
With this notation, the problem to be solved here can be rewritten in a compact form as
\begin{equation}
 \begin{aligned}
 \max_{\bm C \in \mathbb C^{MN\times MN}} & \quad \frac{\frac{W}{N} \log_2 \det \left( \bm I_{KN} + \bm F \bm C \right)}{\frac{1}{\eta N} \trace \bm C +\omega} \\
 \text{s.t.} & \quad \bm C \succeq 0 \\
 & \quad \trace (\bm E_\ell \bm C ) \leq a_\ell, \quad \ell=1,\ldots,U.
 \end{aligned} \label{sub-prob_C}
\end{equation}
This is a fractional program, and the Dinkelbach's algorithm is one of the most widely used tools to solve it~\cite{Jagannathan_1966, Dinkelbach_1967, Crouzeix_1991}. Moreover, since the numerator is concave, and the denominator and the constrains are convex (the denominator is affine), each sub-problem in the Dinkelbach's algorithmin is convex and can therefore be solved in polynomial-time complexity~\cite{Nocedal_2006}. In the following, we first give a brief description of the general Dinkelbach's algorithm, and then we specialize it to Problem~\eqref{sub-prob_C}.

Let $\mathcal S \subseteq R^n$ and $f,h:\mathcal S \rightarrow \mathbb R$; then, a fractional program is the optimization problem
\begin{equation}
 \max_{\bm x \in \mathcal S} \frac{f(\bm x)}{h(\bm x)}. \label{frac_prog_p}
\end{equation}
If $\mathcal S$ is nonempty and compact, $f,h$ are continuous, and $h$ is strictly positive, then the previous problem can be related to
\begin{equation}
 \max_{\bm x \in \mathcal S} \big\{ f(\bm x)- \lambda h(\bm x) \big\} \label{param_prog_p}
\end{equation}
where $\lambda \in\mathbb R$ is a parameter~\cite{Jagannathan_1966, Dinkelbach_1967}. In particular, it can be seen that $\bm x^*$ is optimal for~\eqref{frac_prog_p} if and only if it is optimal for~\eqref{param_prog_p} with $\lambda=\lambda^*$, where $\lambda^*$ is the unique zero of $F(\lambda)=\max_{\bm x \in \mathcal S} \big\{ f(\bm x)- \lambda h(\bm x) \big\}$, and $\lambda^*=f(\bm x^*)/h(\bm x^*)$. Furthermore, $F(\lambda)$ is continuous, convex, and strictly decreasing on $\mathbb R$, $F(\lambda)>0$ for $\lambda<\lambda^*$, and $F(\lambda)<0$ for $\lambda>\lambda^*$. An iterative procedure to compute $\lambda^*$ has been proposed by Dinkelbach~\cite{Dinkelbach_1967} and is reported in Algorithm~\ref{alg_Dinkelbach}. The sequence $\{\lambda_i\}_{i=1}^\infty$ in Dinkelbach' algorithm is strictly increasing and converges to $\lambda^*$. If~\eqref{frac_prog_p} is a concave-convex fractional program,\footnote{The problem in~\eqref{frac_prog_p} is called a concave-convex fractional program if $\mathcal S$ is convex, $f$ is concave, and $g$ is convex; if $g$ is not affine, than $f$ is also required to be nonnegative.} then~\eqref{param_prog_p} is a convex program for $\lambda\geq0$, and Dinkelbach's algorithm converges superlinearly and often (locally) quadratically~\cite{Schaible_1976}.

\begin{algorithm}[t]
\caption{[Dinkelbach] Solution to Problem~\eqref{param_prog_p}}
\label{alg_Dinkelbach}
 \begin{algorithmic}
 \STATE $i=0$
 \STATE choose $\bm x_0\in\mathcal S$
 \REPEAT
 \STATE $i=i+1$
 \STATE $\lambda_i=\frac{f(\bm x_{i-1})}{h(\bm x_{i-1})}$
 \STATE $\bm x_i=\argmax_{\bm x \in \mathcal S} \big\{ f(\bm x)- \lambda_i h(\bm x) \big\}$
 \UNTIL $f(\bm x_i)-\lambda_i h(\bm x_i)$ is sufficiently small
 \end{algorithmic}
\end{algorithm}

Therefore, in order to tackle Problem~\eqref{sub-prob_C} with Dinkelbach's algorithm, we need to solve
\begin{equation}
 \begin{aligned}
 \max_{\bm C \in \mathbb C^{MN\times MN}} & \quad \left\{ \ln \det \left( \bm I_{KN} + \bm F \bm C \right) - \frac{\lambda}{\eta W\log_2\e} \trace \bm C \right\}\\
 \text{s.t.} & \quad \bm C \succeq 0 \\
 & \quad \trace (\bm E_\ell \bm C ) \leq a_\ell, \quad \ell=1,\ldots,U
 \end{aligned} \label{sub-prob_Dinkelbach}
\end{equation}
where we have dropped irrelevant constants. Problem~\eqref{sub-prob_Dinkelbach} is convex and can be managed with standard methods, at least when $MN$ is small.\footnote{Recall that the number of variables of Problem~\eqref{sub-prob_Dinkelbach} is $M^2N^2$ (the real and imaginary parts of the entries that define $\bm C$), and, therefore, even a modest value of $MN$ can be too large to be handled with direct application of standard convex optimization methods.} When instead $MN$ is large, specific methods should be used, so as to reduce the complexity and allow a real-time implementation. In the following, we address this latter case, since $N$ (the length of the STC at the communication system and the number of range bins of the surveillance radar) is typically large.

First notice that we can restrict our attention to the case\footnote{From~\eqref{rho_max}, the initial covariance matrix in the block coordinate ascent algorithm can be chosen such that $\text{EE}>0$, and EE is non-decreasing with the iteration number. This implies that $\lambda_1>0$, since the initial point in Algorithm~\ref{alg_Dinkelbach} can be chosen to be the covariance matrix at the previous iteration of the block coordinate ascent algorithm. Finally, since  and $\{\lambda_i\}_{i=1}^\infty$ is stricltly increasing, $\lambda_i>0$, for all $i$.} $\lambda>0$. Then, if Problem~\eqref{sub-prob_C} is strictly feasible (and this holds true if there exists a Hermitian positive definite matrix satisfying the linear matrix inequality constraints~\cite{Boyd_Vandenberghe_2004}), the solution to Problem~\eqref{sub-prob_Dinkelbach} is
\begin{equation}
 \bm C= \bm Z_{\bm \mu}^{-1}  \bm V_{\bm \mu} \left(\bm I_\Delta -\bm \Xi_{\bm \mu}^{-1} \right)_+ \bm V_{\bm \mu}\herm \bm Z_{\bm \mu}^{-1}\label{sol_C}
\end{equation}
where
\begin{equation}
 \bm Z_{\bm \mu} = \left( \frac{\lambda}{\eta W \log_2\e} \bm I_{MN} +\sum_{\ell=1}^U \mu_\ell \bm E_\ell \right)^{1/2} \label{Z_mat}
\end{equation}
$\bm \Xi_{\bm \mu}= \diag (\xi_{\bm \mu, 1}, \ldots, \xi_{\bm \mu, \Delta})$ contains the non-zero eigenvalues of $\bm Z_{\bm \mu}^{-1} \bm F \bm Z_{\bm \mu}^{-1}$, with $\Delta\leq N\min\{K,M\}$ denoting the rank of $\bm F$, $\bm V_{\bm \mu} \in \mathbb C^{MN\times \Delta}$ is the matrix containing the corresponding unit-norm eigenvalues, and $\bm \mu = (\mu_1 \; \cdots\; \mu_U)\transp$ solves the dual problem
\begin{equation}
 \begin{aligned}
 \max_{\bm \mu \in\mathbb R^U} & \quad g(\bm \mu) \\
 \text{s.t.} & \quad \mu_\ell\geq 0, \quad \ell=1,\ldots,U.\label{dual_prob}
 \end{aligned}
\end{equation}
with
\begin{equation}
 g(\bm \mu) =\sum_{\substack{\ell=1\\\ell: \,\xi_{\bm \mu, \ell}>1}}^\Delta \left( 1-\frac{1}{\xi_{\bm \mu, \ell}}  - \ln \xi_{\bm \mu, \ell} \right) -  \sum_{\ell=1}^U \mu_\ell a_\ell \label{dual_fun}
\end{equation}
being the dual objective function. This generalizes the result in~\cite{Zhang_2010}, derived for the rate optimzation problem, and the proof is reported in the Appendix. Notice that $\bm Z_{\bm \mu}$ is positive definite, since $\lambda>0$ and $\bm E_\ell\succeq0$ for $\ell=1,\ldots,U$, and, unless all the SDR constraints are inactive, it is not a scaled identity matrix, so that the eigenvectors of~\eqref{sol_C} are in general not matched to those of the equivalent channel matrix $\bm F$.

The dual problem is simpler than Problem~\eqref{sub-prob_Dinkelbach}, since it has no matrix inequality or matrix variable, and the number of (scalar) variables is equal to the number of inequality constraints in the original problem. Several methods can be used to solve Problem~\eqref{dual_prob}. One is the projected gradient method: at each step, the algorithm selects the next point by moving towards the gradient direction (like the gradient method) and then projecting onto the feasible set; in the present case, the projection operation is just the positive part of each coordinate of the point. The iterative step of the algorithm is, therefore,
\begin{equation}
 \mu_\ell^{(k+1)}= \left( \mu_\ell^{(k)} +\alpha^{(k)} \frac{\partial g(\bm \mu^{(k)})}{\partial \mu_\ell} \right)^+ \label{projected_gradient_update}
 \end{equation}
where $\alpha^{(k)}$ is the step-size parameter\footnote{In order to reduce complexity, the step-size can be fixed ahead in time (and not selected via a line search), as in the sub-gradient method; in this case, many choices are available~\cite{Shor_1985}, and a simple one that guarantees convergence is $\alpha^{(k)}$ such that $\alpha^{(k)}\geq 0$, $\lim_{k\rightarrow \infty} \alpha^{(k)}=0$, and $\sum_{k=1}^\infty \alpha^{(k)}=\infty$ (e.g., a universal choice is $\alpha^{(k)}=1/k$).} at the $k$-th iteration, and\footnote{In this case~\cite[cfr.~Th~6.3.3]{Bazaraa_2006}, the partial derivatives coincide with the residuals for the constraints in the primal problem in~\eqref{sub-prob_Dinkelbach} with $\bm C$ as in~\eqref{sol_C}.}
\begin{equation}
\frac{\partial g ( \bm \mu)}{\partial \mu_\ell} =\trace\left( \bm E_\ell \bm Z_{\bm \mu}^{-1}  \bm V_{\bm \mu} \left(\bm I_\Delta -\bm \Xi_{\bm \mu}^{-1} \right)_+ \bm V_{\bm \mu}\herm \bm Z_{\bm \mu}^{-1} \right) -a_\ell.
\end{equation}

The projected gradient method is characterized by a slow convergence rate. If a faster convergence is needed (at the price, however, of an increased computational complexity), we can resort to the quasi-Newton projected gradient method presented in~\cite{Bertsekas_1982, Kim_2010, Schmidt_2012}. At each iteration, the variables are partitioned in two groups, named \emph{restricted} and \emph{free}. The set of restricted variables is
\begin{equation}
 \mathcal R_k =\left\{ \ell\in\{1,\ldots,U\}: \mu^{(k)}_\ell \leq \epsilon \text{ and } \frac{\partial g ( \bm \mu^{(k)})}{\partial \mu_\ell} <0 \right\}
\end{equation}
where $\epsilon$ is a small positive constant (cfr.~\cite{Bertsekas_1982}). This set contains the variables that are approximately zero, and for which the objective function can be increased by moving towards negative values (i.e., outside the boundary): these variables are kept fixed in the $k$-th iteration. The set of free variables, denoted by $\mathcal F_k$, is simply the complement of $\mathcal R_k$ in $\{1,\ldots,U\}$: these variables are updated through a projected quasi-Newton step. Let $\bm S^{(k)}\in\mathbb R^{U\times U}$ be a symmetric positive definite matrix, that is an approximation of the inverse of the Hessian matrix;\footnote{It can be shown that the objective function of Problem~\eqref{dual_prob} is differentiable but not twice differentiable.} in order to limit the complexity, $\bm S^{(k)}$ can be computed starting from $\bm S^{(k-1)}$, and a popular choice is the Broyden-Fletcher-Goldfarb-Shanno (BFGS) update rule or the damped BFGS update rule, so as to guarantee that $\bm S^{(k)}$ be positive definite~\cite{Nocedal_2006}. The $k$-th iteration of the algorithm is, therefore,
\begin{equation}
\begin{cases}
 \mu_\ell^{(k+1)} = \left( \mu_\ell^{(k)} +\alpha^{(k)} \sum_{i \in \mathcal F_k} S^{(k)}_{\ell,i} \frac{\partial g(\bm \mu^{(k)})}{\partial \mu_i} \right)^+ ,&\text{if } \ell \in \mathcal F_k\\
 \mu_\ell^{(k+1)} =\mu_\ell^{(k)}, &\text{if } \ell \in \mathcal R_k
\end{cases}\label{quasi-Newton_projected_gradient_update}
\end{equation}
where $S^{(k)}_{\ell,i}$ is the element $(\ell,i)$ of $\bm S^{(k)}$, and the step-size $\alpha^{(k)}$ is computed though a backtracking (over the free variables only) Armijo line search~\cite{Bertsekas_1982, Kim_2010}. This algorithm has been shown to be globally convergent, and, under certain conditions, it achieves local superlinear convergence~\cite{Bertsekas_1982, Kim_2010, Schmidt_2012}.

\subsection{The complete algorithm} \label{algorithm_sec}

The block coordinate ascent method, used to find a sub-optimum solution to Problem~\eqref{opt_prob}, and the Dinkelbach's routine with the projected gradient (or the quasi-Newton projected gradient) algorithm, used to solve the communication codebook optimization problem in Sec.~\ref{comm_syst_sol}, are integrated in Algorithm~\ref{alg_2}. Loops end when a sufficiently small increase of the objective function in two consecutive iterations is observed or when a specified maximum number of iterations is reached.

\begin{algorithm}[t]
\caption{Sub-optimum solution to Problem~\eqref{opt_prob}}
\label{alg_2}
 \begin{algorithmic}
 \STATE chose $\{\bm w_{n,j}\}_{(n,j)\in\mathcal X}$, $P_r$, $\bm C$ satisfying the constraints
 \REPEAT
 \STATE update $\{\bm w_{n,j}\}_{(n,j)\in\mathcal X}$ with~\eqref{w_update}
 \STATE update $P_r$ with~\eqref{Pr_update}
 \STATE evaluate $\bm F$ in~\eqref{mat_F}
 \REPEAT
 \STATE $\lambda=\frac{\frac{W}{N}\log_2 \det \left( \bm I_{KN} + \bm F \bm C \bm F\herm \right)}{\frac{1}{\eta N} \trace \bm C +\omega}$
 \STATE chose $\bm \mu\in\mathbb R_+^U$
 \STATE update $\bm C$ with~\eqref{sol_C}
 \REPEAT
 \STATE update $\bm \mu$ with~\eqref{projected_gradient_update} or with~\eqref{quasi-Newton_projected_gradient_update}
 \STATE update $\bm C$ with~\eqref{sol_C}
 \UNTIL{convergence}
 \UNTIL {$\ln \det ( \bm I_{KN} + \bm F \bm C \bm F\herm )- \frac{\lambda}{\eta W\log_2\e} \trace \bm C$ is sufficiently small}
 \UNTIL{convergence}
 \RETURN $\{\bm w_{n,j}\}_{(n,j)\in\mathcal X}$, $P_r$, $\bm C$
 \end{algorithmic}
\end{algorithm}

Let us examine the complexity of Algorithm~\ref{alg_2}, and assume, at first, that the projected gradient algorithm is used. The computational complexity of the innermost loop is dominated by the cost of the update of $\bm C$ in~\eqref{sol_C}, that requires evaluating the matrix $\bm Z_{\bm \mu}$ in~\eqref{Z_mat} and the eigenvalue decomposition of $\bm Z_{\bm \mu}^{-1} \bm F \bm Z_{\bm \mu}^{-1}$. The cost for evaluating $\bm Z_{\bm \mu}$ is $\mathcal O(N^2M^2 U+ N^3M^3)$, due to the matrix summation and to the square root matrix operation, while the cost for the eigenvalue decomposition is $\mathcal O(N^3M^3)$, so that the overall cost of the innermost loop is $\mathcal O\big(N^2M^2\max\{MN,U\}\big)$. The remaining operations in the middle loop have a computational complexity $\mathcal O\big(N^3K \max\{M^2,K^2\}\big)$ (ruled by the matrix multiplication and the determinant in the update of $\lambda$), so that the overall cost of Dinkelbach's algorithm is $\mathcal O\big(N^2\max\{NM^3, NK^3, M^2U\}\big)$. The computational complexity of the remaining operations in the outermost loop is dictated by the radar filters and power update in~\eqref{w_update} and~\eqref{Pr_update}, respectively, and by the evaluation of the matrix $\bm F$ in~\eqref{mat_F}: the first two operations have a cost of $\mathcal O(N^3J)$, while the third has a cost of $\mathcal O\big(N^3K^2\max\{M,K\}\big)$, so that the overall cost is $\mathcal O\big(N^3\max\{J,K^2M,K^3\}\big)$. In conclusion, the computational complexity of Algorithm~\ref{alg_2} is $\mathcal O\big(N^2\max\{NM^3, NK^3, NJ, M^2U\}\big)$, that is at most $\mathcal O\big(N^3\max\{K^3,M^3,M^2J\}\big)$, since $U\leq (N-L+1)J+1$.

If the quasi-Newton projected gradient algorithm is instead used, the computational complexity of the innermost loop is ruled by the update of $\bm \mu$ in~\eqref{quasi-Newton_projected_gradient_update}. Since the step-size selection through the Armijo line search does not increase the complexity (it requires the computation of $g(\bm \mu)$, that has the same complexity as the update of $\bm C$), the only additional cost with respect to the projected gradient algorithm is the update of the gradient-scaling matrix $\bm S^{(k)}$ that, with the BFGS routine, has a cost $\mathcal O (U^2)$. In this case, the computational complexity of Algorithm~\ref{alg_2} is $\mathcal O\big(N^2\max\{NK^3,NM^3,NJ,M^2U^2\}\big)$.

The proposed algorithm runs in a fusion center, that can possibly be one of the two coexisting systems. The additional (with respect to the non-shared spectrum case) amount of cognition and/or cooperation required is quite limited. Indeed, when the spectrum is not shared, the second order statistic of the noise (i.e., its power $P_v$) and the instantaneous channel realization $\bm H$ must be estimated at the communication receiver, e.g., through periodic training phases; then the covariance matrix $\bm C$ (or channel realization $\bm H$, if the transmitter is in charge of finding the optimum covariance matrix) must be sent through a feedback channel to the transmitter. Similarly, the second order statistics of the disturbance, i.e., the noise power $P_u$ and the clutter strengths $\{\sigma^2_{\gamma, i, j}\}$, must be periodically estimated/updated at the radar receiver and sent to the transmitter (if they are not colocated) so as to adjust the power level and obtain the required detection/estimation performance. In the present coexisting scenario, instead, the additional parameters that should be estimated are the mutual delay between the radar and communication transmitters, $\nu_0$, and the second order statistics of the mutual interference, i.e., $\{\bm \Sigma_{\alpha, i}\}$ and $\{\sigma^2_{\beta,m,m',i,j}\}$, that usually vary on a time scale smaller than that of the instantaneous channel matrix $\bm H$. The complete list of the steps that must be taken at regular time intervals is listed below.
\begin{enumerate}
 \item The radar transmitter is idle, while the communication transmitter sends training data; the communication receiver estimates $\bm H$ and $P_v$, while the radar receiver estimates $\{\sigma^2_{\beta,m,m',i,j}\}$, and $\nu_0$.
 \item The communication transmitter becomes idle and the radar active; the radar receiver estimates $\{\sigma^2_{\gamma, i, j}\}$ and $P_u$, while the communication receiver estimates $\{\bm \Sigma_{\alpha, i}\}$.
 \item The fusion center runs Alg.~\ref{alg_2} and sends $\bm C$ to the communication transmitter, $P_r$ to the radar transmitter, and $\{\bm w_{i,j}\}$ to the radar receiver. 
\end{enumerate}

\section{Numerical examples}\label{num_ex_sec}

Here we examine two systems operating at 4~GHz over a bandwidth of $W=1$~MHz. The range resolution (dictated by $W$) is 150~m, and the PRF is 10~kHz (i.e., $T=100$~$\mu$s);\footnote{We underline that these are illustrative values, that, however, are compatible with the typical values of a pulse Doppler radar~\cite{Skolnik_2001} and with the range of time scales over which radar signal processing operations typically take place~\cite{Richards_2005}.} this implies that $N=100$ range cells are defined, with maximum non-ambiguous range equal to 15~km. The pulses are modulated by a Barker code of length $L=5$. The maximum peak power is $P_{r,\text{max}}^\text{peak}=500$~W, so that the maximum average power is $P_{r,\text{max}}=P_{r,\text{max}}^\text{peak}L/N=25$~W. At the receiver side, $J=3$ orthogonal beams are formed. The power spectral density (PSD) of the noise is $\sigma^2_u= 4\times 10^{-21}$~W/Hz, so that $P_u=F\sigma^2_u W=2.39\times 10^{-14}$~W, where $F=6$~dB is the receiver noise figure. We test the system for a specified signal level, that corresponds to a target with a radar cross-section that increases with the distance from the radar: specifically, $\sigma^2_{g,n,j}=4.8\times10^{-16}$, for all $(n,j)\in\mathcal X$, so that $\sigma^2_{g,n,j} NP_{r,\text{max}} /P_u= 17$~dB (this is the largest achievable signal-to-noise ratio). As to the clutter, we set $\sigma^2_{\gamma,i,j}=4.8\times10^{-17}$, for all $(n,j)\in\mathcal X$, so that $\sigma^2_{\gamma,i,j} NP_{r,\text{max}} /P_u= 7$~dB (this is the largest possible clutter-to-noise ratio in each resolution cell). Also, we require the same minimum SDR in any radar resolution cell, i.e., $\rho_{i,j}=\rho$, for all $(i,j)\in\mathcal X$. The communication transmitter has $M=2$ antennas, with maximum average transmit power $P_{c,\text{max}}=10$~mW, power amplifier efficiency $\eta=0.85$, and circuit power required to operate the link $\omega=10$~mW; the receiver is located at a distance of 100~m and is equipped with $K=2$ receive antennas. The PSD of the noise is $\sigma^2_v= 4\times 10^{-21}$~W/Hz, and $P_v=F\sigma^2_v W=2.39\times 10^{-14}$~W. The entries of the channel matrix $\bm H$ are generated following a CCSG distribution with variance $\sigma^2_h=3\times 10^{-10}$, so that $\sigma^2_h P_{c,\text{max}} / P_v=21$~dB (this is the largest achievable signal-to-noise ratio).

As to the mutual interference between the two systems, we set $\bm \Sigma_{\alpha,i}=\sigma^2_{\alpha,i}\bm I_K$, and $\sigma^2_{\beta,m,m',i,j}=\sigma^2_{\beta,i,j}$, if $m=m'$, and $\sigma^2_{\beta,m,m',i,j}=0$, otherwise: this models the case where no line-of-sight component is present, and independent rays from the rich scattering environment arrive at both receivers. The coefficients $\sigma^2_{\alpha,i}$ and $\sigma^2_{\beta,i,j}$ are set equal to either 0 or $\sigma^2$, and we test $\sigma^2=1.2\times10^{-13}$ and $\sigma^2=1.2\times10^{-11}$. The former value corresponds to a lightly interferenced scenario, and results in $\sigma^2_{g,n,j} P_{r,\text{max}}/(\sigma^2P_{c,\text{max}}) = 10$ dB at the radar side and $\sigma^2_h P_{c,\text{max}}/(\sigma^2P_{r,\text{max}})=0$ dB at the communication system side; the latter value, instead, describes a strongly interferenced situation, where we have $\sigma^2_{g,n,j} P_{r,\text{max}}/(\sigma^2P_{c,\text{max}}) = -10$ dB at the radar and $\sigma^2_h P_{c,\text{max}}/(\sigma^2P_{r,\text{max}})=-20$ dB at the communication system. The fraction of non-zero entries in $\{\sigma^2_{\alpha,i}\}_{i=0}^{N-1}$ and $\{\sigma^2_{\beta,i,j}\}_{i=0}^{N-1}$, for $j=1,\ldots,J$, is denoted $\delta$, and different values are tested: smaller $\delta$'s correspond to scenarios with a limited number of scatterers, while larger $\delta$'s refer to a \emph{crowded} environments.

In this framework, we use Algorithm~\ref{alg_2} with the quasi-Newton projected gradient method for the $\bm \mu$-update to suboptimally solve the proposed joint design problem in~\eqref{opt_prob}, and we compare its performance with that of two other cases. In the first one, we consider \emph{non-interfering} systems, and Problem~\eqref{opt_prob} is solved when $\delta = 0$ and/or $\sigma^2=0$. It can easily be shown that, at the radar side, a solution is
\begin{subequations}
\begin{align}
 P_r&=P_{r,\text{max}}\\
 \bm w_{n,j}&=\left(P_{r,\text{max}}\sum_{i=0}^{N-1} \sigma^2_{\gamma,i,j} \bm q_i \bm q_i\herm +P_u\bm I_N\right)^{-1} \bm q_n, \; (n,j)\in\mathcal X
\end{align}\label{radar_only}%
\end{subequations}
while, at the communication system side, $\bm C$ can be found with Dinkelbach's algorithm, where, at each iteration, the solution to the problem parametrized by $\lambda>0$ is given by standard waterfilling over the channel $\bm r = (\bm H \otimes \bm I_N) \bm c+ \bm v$ with power constraint
\begin{equation}
\min \left\{P_{c,\text{max}}, N \sum_{i=1}^{\theta} \left(\frac{\eta W \log_2\e}{\lambda}-\frac{1}{\zeta_i^2}\right)^+\right\}
\end{equation}
with $\theta$ denoting the rank of $\bm H$ and $\{\zeta_1,\ldots,\zeta_\theta\}$ its non-zero singular values. This represents an upper bound to the system performance. In the second case, the previous solution is incorrectly used when the mutual interference is present; this corresponds to the case where each system independently maximizes its own performance measure (EE and minimum SDR over the radar resolution cells) ignoring the presence of the other system, and is therefore referred to as \emph{disjoint design}. This is the lower bound of non-cooperative systems.

The performance is evaluated by averaging 200 Monte Carlo runs, where the channel matrix $\bm H$, the indexes of the non-zero entries in $\{\sigma^2_{\alpha,i}\}_{i=0}^{N-1}$ and $\{\sigma^2_{\beta,i,j}\}_{i=0}^{N-1}$, and the delay $\nu_0$ of the communication transmitter with respect to the radar transmitter are randomly generated in each run. This can effectively model the situation were the mutual position of the two systems and the position of the surrounding scatterers may change with time, as may the mutual transmit delay (e.g., because one or both systems are either idle or performing different tasks for some periods of time).

\begin{figure}[t]
 \centering
 \centerline{\includegraphics[width=1.05\columnwidth]{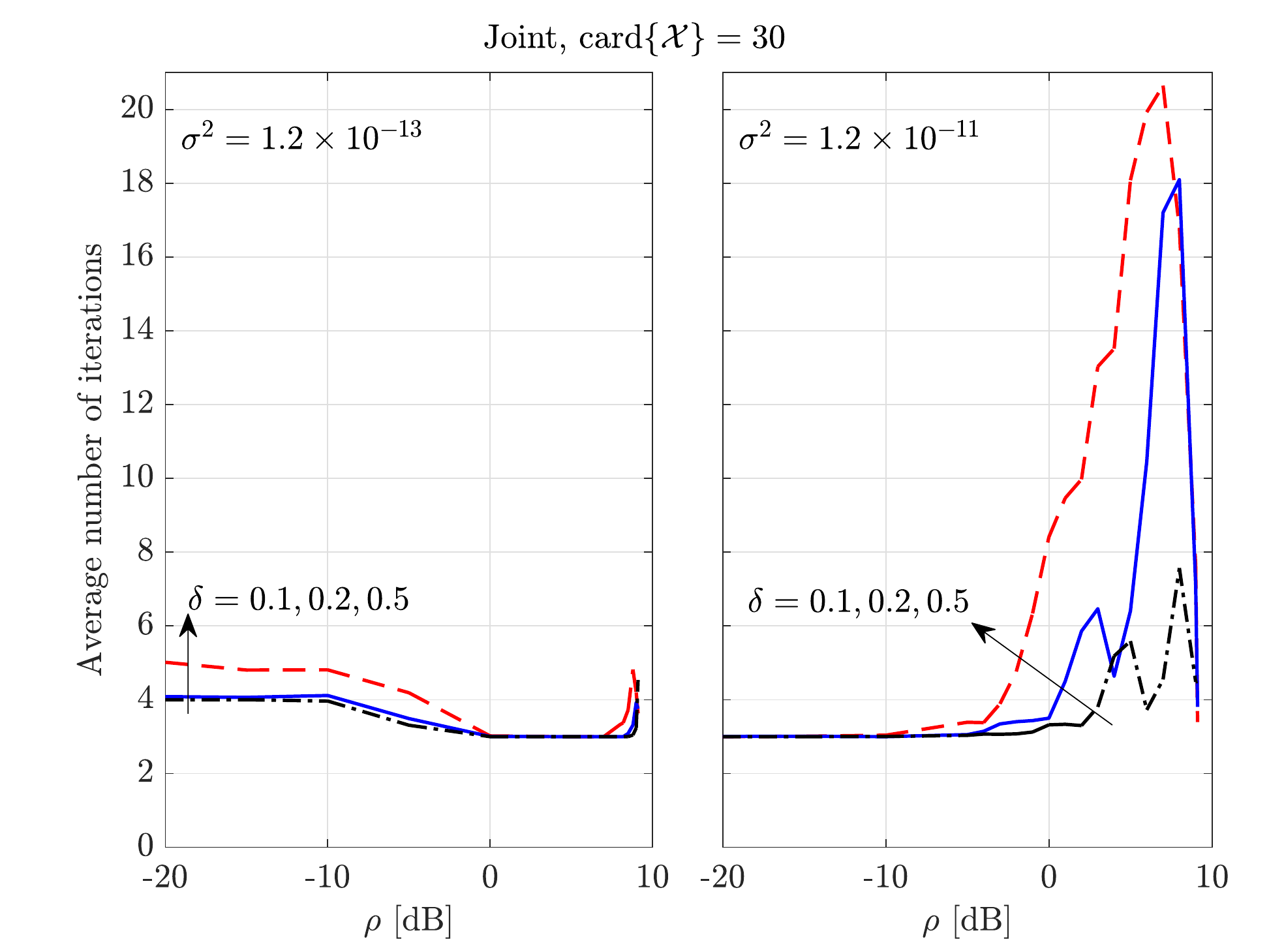}}
\caption{Average number of iterations (among the Monte Carlo runs) needed to observe a relative improvement in the EE smaller than $10^{-5}$ in the block coordinate ascent algorithm of the joint design strategy versus the minimum required SDR for different values of  intensity of the interference, $\sigma^2$, and density of the interference scatterers, $\delta$, when 30 resolution cells are protected.} \label{fig_01}
\end{figure}

Fig.~\ref{fig_01} reports the average number of iterations needed to observe a relative improvement in the EE smaller than $10^{-5}$ in the 200 runs of the block coordinate ascent algorithm (outermost loop of Algorithm~\ref{alg_2}). The number of iterations is shown versus the minimum required SDR at the radar, $\rho$, for different values of the density of the interference scatterers, $\delta$, and intensity of the interference, $\sigma^2$, when the number of protected radar resolution cells is $\card \mathcal X =30$. It can be seen that the algorithm rapidly converges in all inspected cases, occasionally requiring a larger average number of iterations only when there is a strong interference and a high performance is requested at the radar side (i.e., large values of $\rho$).

\begin{figure}[t]
 \centering
 \centerline{\includegraphics[width=1.08\columnwidth]{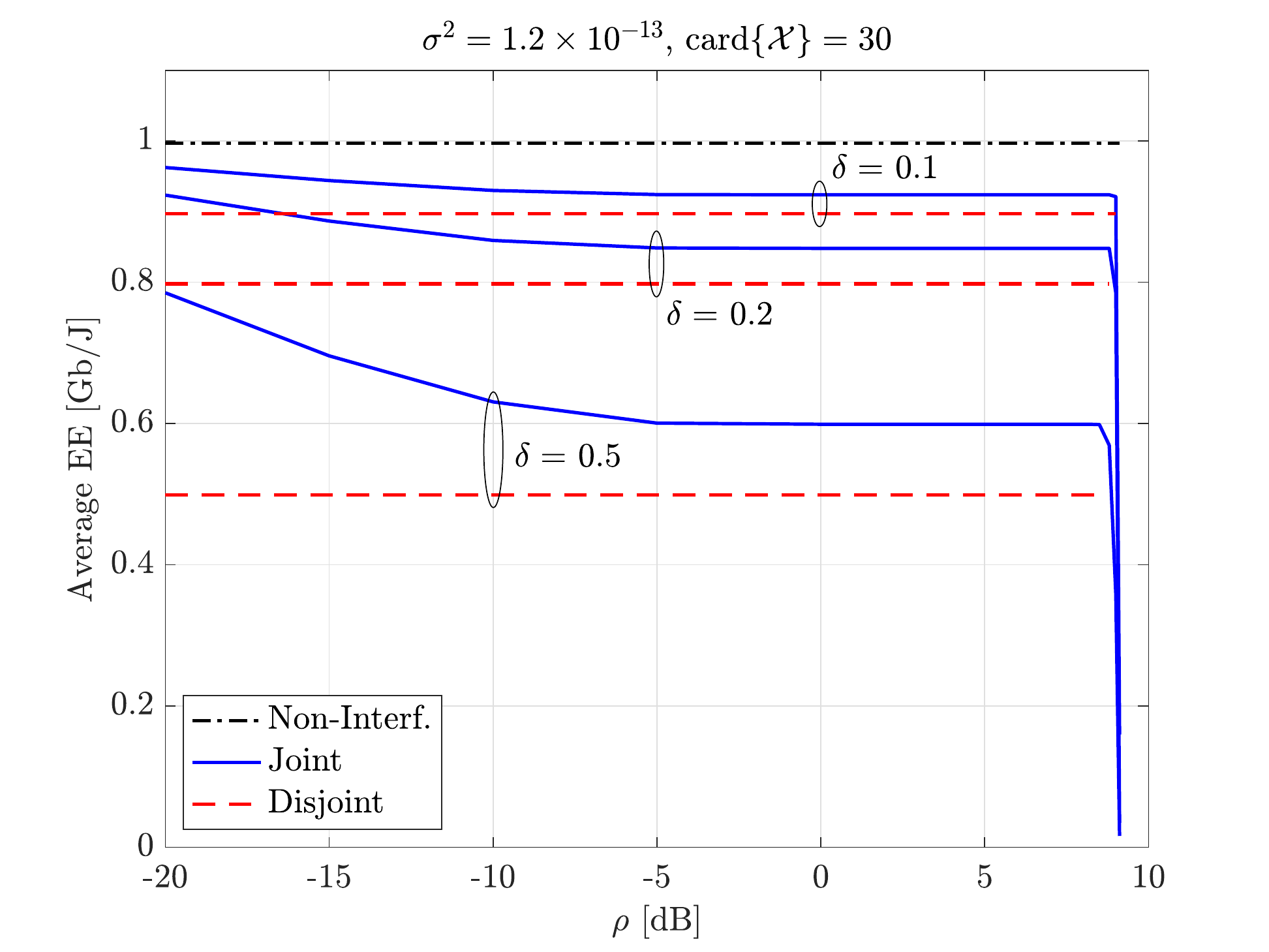}}
\caption{Average energy efficiency versus the minimum required SDR for a joint and disjoint design and for non-interfering systems, and for different values of the density of the interference scatterers, $\delta$, when the intensity of the interference is $\sigma^2=1.2\times 10^{-13}$, and 30 resolution cells are protected; for comparison purposes, the case of non-interfering systems is also included.} \label{fig_02}
\end{figure}

Fig.~\ref{fig_02} shows the average energy efficiency versus $\rho$ for different values of $\delta$, when $\sigma^2=1.2\times 10^{-13}$ and $\card \mathcal X =30$. Notice that, not all $\rho$'s are allowed (as it can also be seen in Fig.~\ref{fig_01}): in fact, from~\eqref{rho_max}, Problem~\eqref{opt_prob} admits a solution only if $\rho\leq 9.2$~dB, and such SDR can be achieved when there is no interference from the communication system, and the disturbance is only due to noise and clutter. Notice also that the curves corresponding to the non-interfering and disjoint cases are half-lines with zero slope, since we have the same EE for all feasible $\rho$'s: in the non-interfering case, the feasible set is $\rho \leq 9.2$~dB, while, in the disjoint design, it is $\rho\leq \rho^*< 9.2$~dB, since the mutual interference is now present and not accounted for. Furthermore, $\rho^*$ is a random variable, since it depends on the channel realization $\bm H$: namely, if such realization is \emph{good} (\emph{bad}), the communication transmitter needs a small (large) power to implement the disjoint design, and this results in a low (large) interference at the radar, so that the corresponding realization of $\rho^*$ is large (small). In Fig.~\ref{fig_02}, we therefore extend the curves corresponding to the disjoint design up to the largest realization of $\rho^*$ observed in the Monte Carlo runs, but each value of $\rho$ is characterized by the probability that such value is actually feasible (i.e., the probability that $\rho^*\geq \rho$).

\begin{figure}[t]
 \centering
 \centerline{\includegraphics[width=1.08\columnwidth]{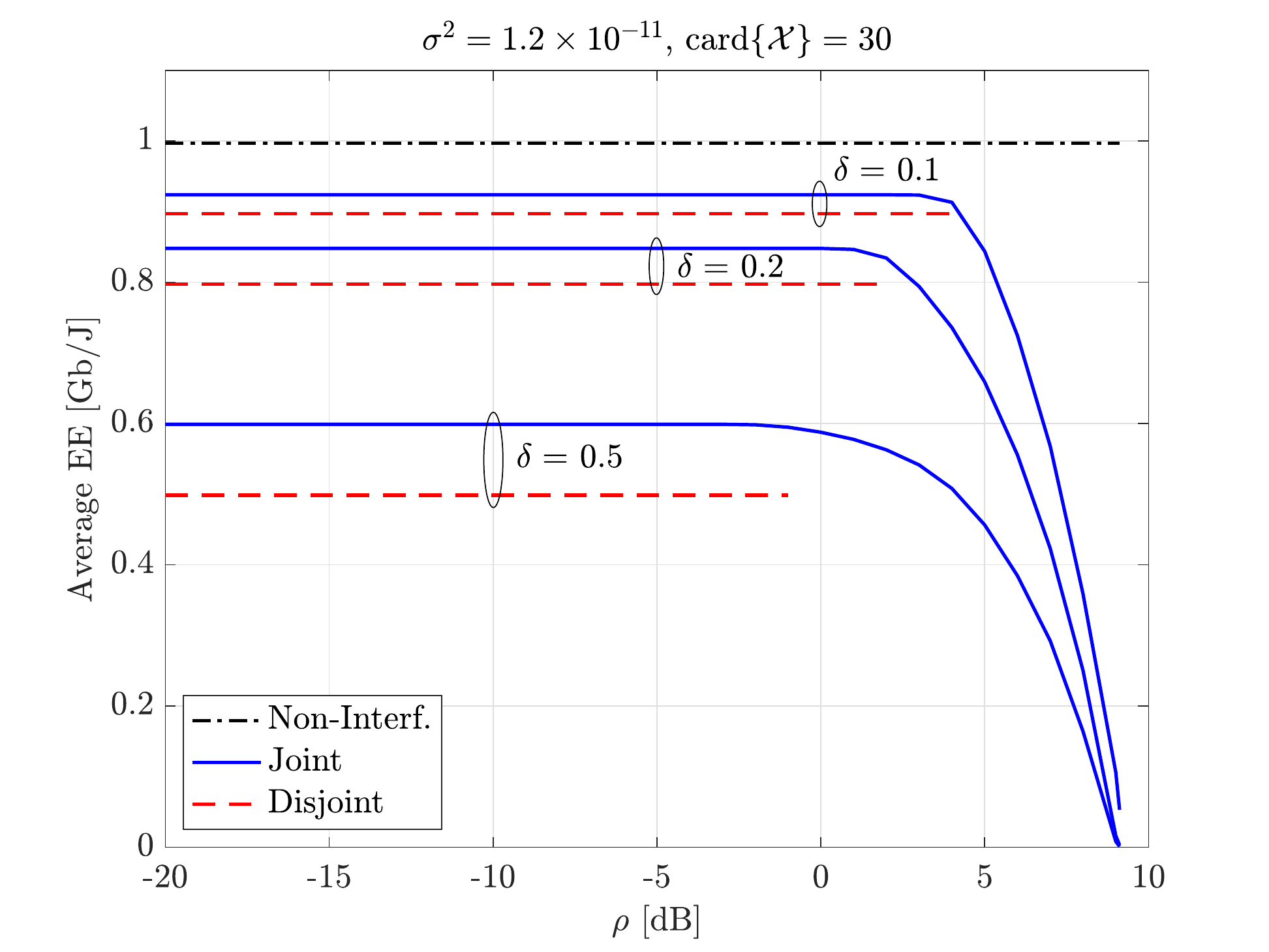}}
\caption{Average energy efficiency versus the minimum required SDR for a joint and disjoint design, and for different values of the density of the interference scatterers, $\delta$, when the intensity of the interference is $\sigma^2=1.2\times 10^{-11}$, and 30 resolution cells are protected; for comparison purposes, the case of non-interfering systems is also included.} \label{fig_03}
\end{figure}

\begin{figure}[t]
 \centering
 \centerline{\includegraphics[width=1.05\columnwidth]{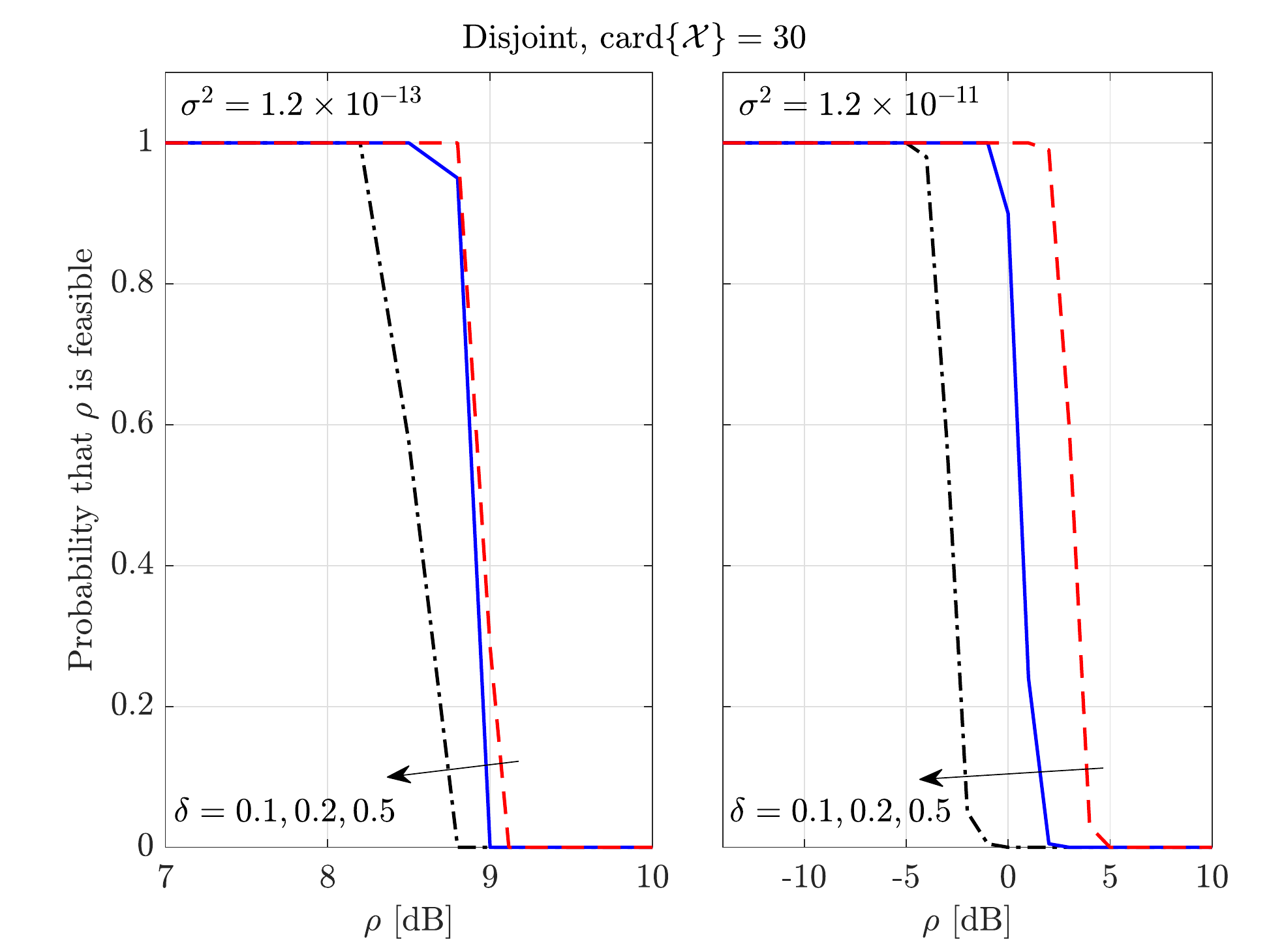}}
\caption{Average energy efficiency versus the minimum required SDR for different design strategies and different values of the intensity of the interference, when the density density of the interference scatterers is $\delta=0.5$, and 30 resolution cells are protected.} \label{fig_04}
\end{figure}

From Fig.~\ref{fig_02}, we see that the EE of the joint design decreases with $\rho$ and $\delta$, and tends to the upper bound of the non-interfering case when $\rho\rightarrow -\infty$~dB. Notice that the joint design outperforms the disjoint one not only in the value of EE, where the gap becomes significant for high $\delta$'s and/or low $\rho$'s, but also in the achievable values of minimum SDR, where the gap is around 0.5--1~dB. The gain of the joint design is much larger when the mutual interference is stronger ($\sigma^2=1.2\times 10^{-11}$), as it can be seen from Fig.~\ref{fig_03}. In this case, the gap in terms of achievable SDR with respect to the disjoint design is significant, and amounts to 5, 7, and 10~dB for $\delta=0.1$, 0.2, and 0.5, respectively. These gaps, however, increase when referred to values of the EE achievable in 100\% of the cases, as it can be seen from Fig.~\ref{fig_04}, that shows the probability that the SDR constraint is feasible for the disjoint design with different values of $\delta$ and $\sigma^2$ when $\card \mathcal X =30$: indeed, if the 9.2~dB limit of the joint design in Figs.~\ref{fig_02} and~\ref{fig_03} is compared with the largest value of $\rho$ achievable by the disjoint design in 100\% of the cases reported in Fig.~\ref{fig_04}, it can be seen that the latter results into a loss of about 1~dB in the lightly interfered case and of as much as 8--14~dB in the strongly interfered case.

\begin{figure}[t]
 \centering
 \centerline{\includegraphics[width=1.08\columnwidth]{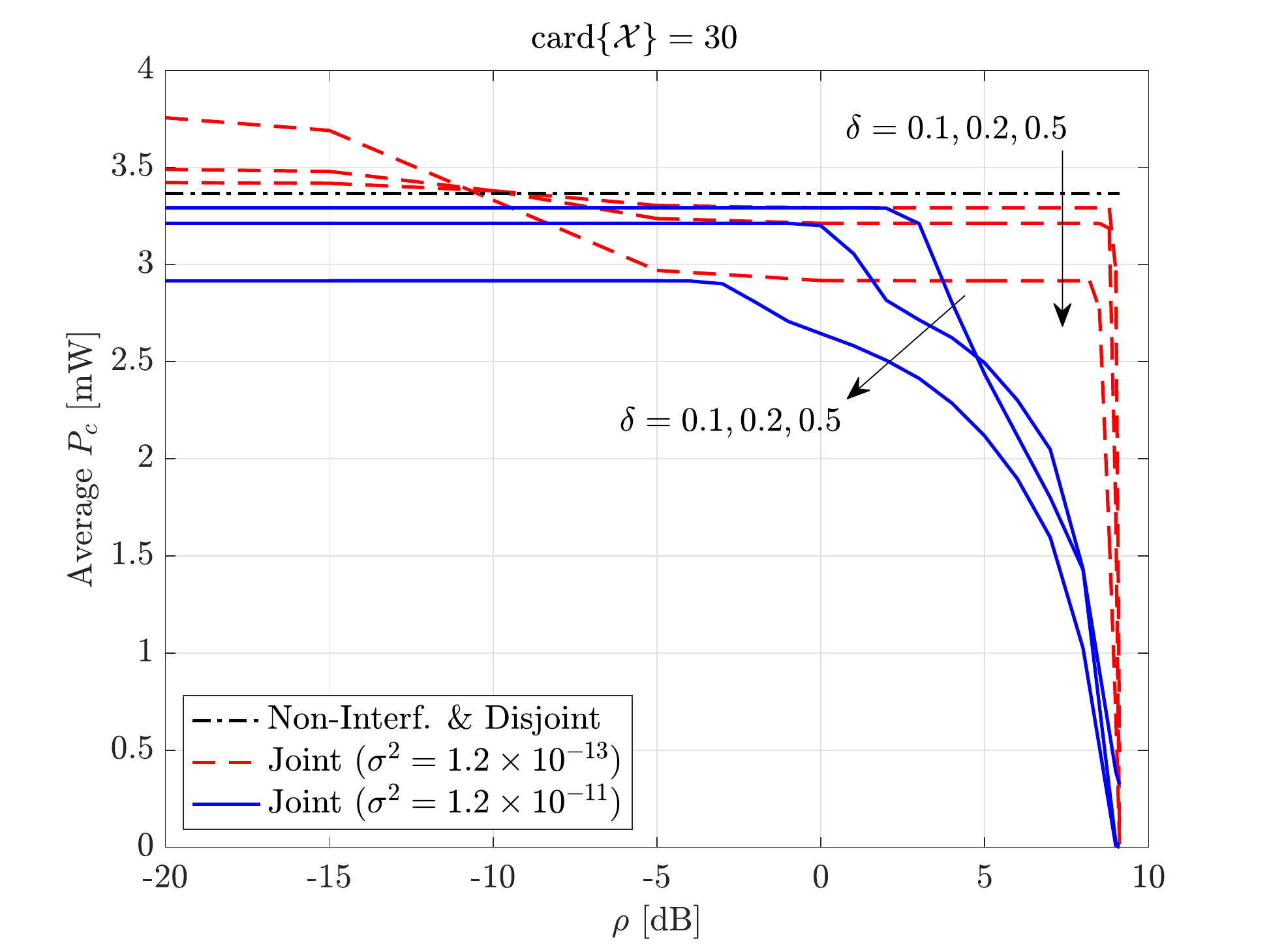}}
\caption{Average communication transmit power versus the minimum required SDR for a joint and disjoint design, different values of the density of the interference scatterers, $\delta$, and intensity of the interference, $\sigma^2$, when 30 resolution cells are protected; for comparison purposes, the case of non-interfering systems is also included.} \label{fig_05}
\end{figure}

\begin{figure}[t]
 \centering
 \centerline{\includegraphics[width=1.08\columnwidth]{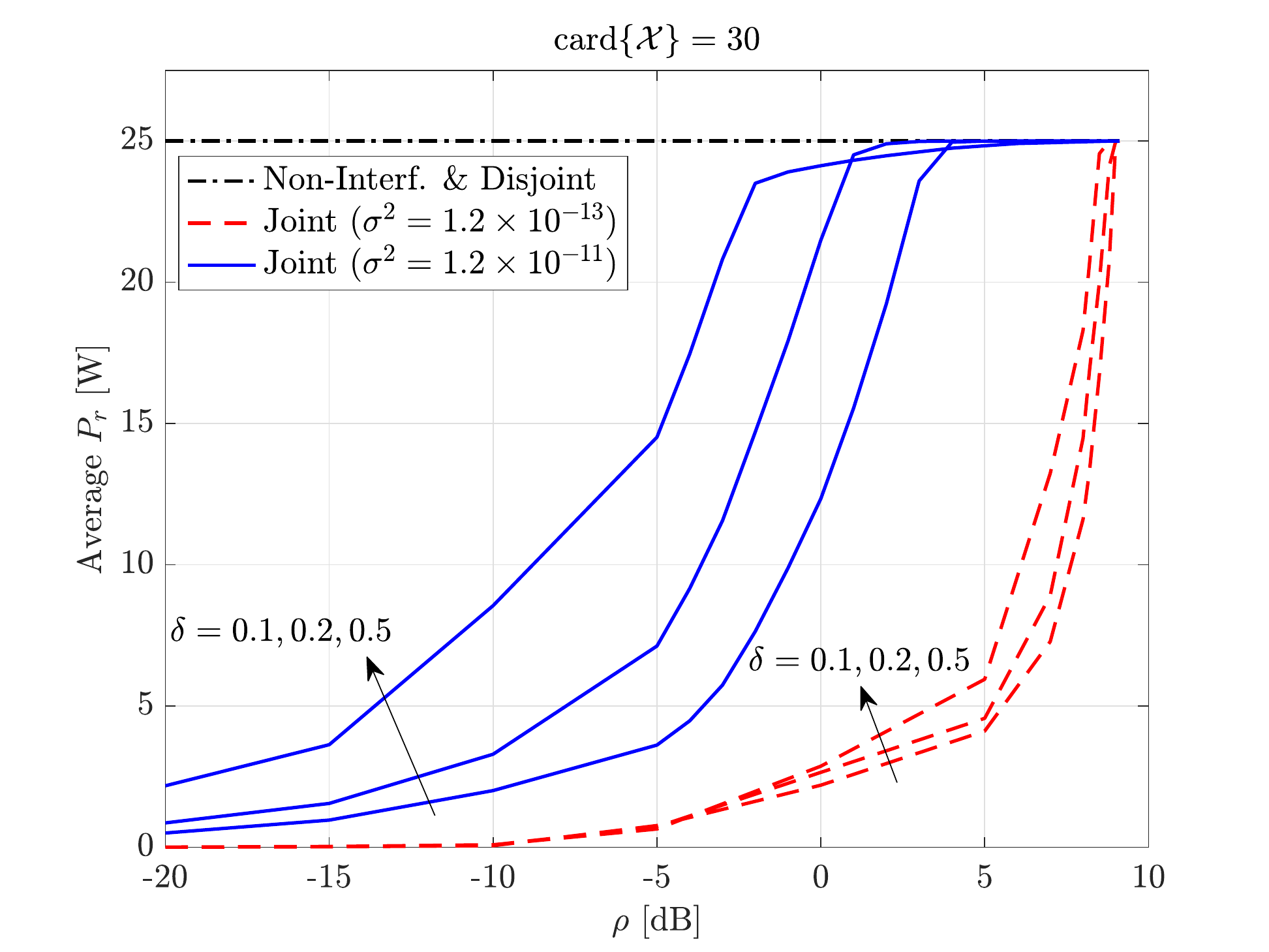}}
\caption{Average radar transmit power versus the minimum required SDR for a joint and disjoint design, different values of the density of the interference scatterers, $\delta$, and intensity of the interference, $\sigma^2$, when 30 resolution cells are protected; for comparison purposes, the case of non-interfering systems is also included.} \label{fig_06}
\end{figure}

Figs.~\ref{fig_05} and~\ref{fig_06} show, as a function $\rho$, the average transmit communication and radar powers, respectively, corresponding to the cases inspected in Figs.~\ref{fig_02} and~\ref{fig_03}. Clearly, when the minimum performance level required at the radar is increased, the interference becomes stronger, or the density of the interference scatterers becomes higher (large values of $\rho$, $\sigma^2$, or $\delta$, respectively), the transmit power must be increased at the radar side. At the communication side, instead, the transmit power must be decreased when $\rho$ is very close to the maximum allowed value: in all other cases, the power level lies around the value of the non-interfering systems. Notice that there is no value of $\rho$ where both systems are transmitting with the power level used in the non-interfering case, this meaning that, in order to mitigate the mutual interference and allow coexistence, space-time beamforming alone is not sufficient, and power control is also needed.

\begin{figure}[t]
 \centering
 \hspace{-10pt}\centerline{\includegraphics[width=1.02\columnwidth]{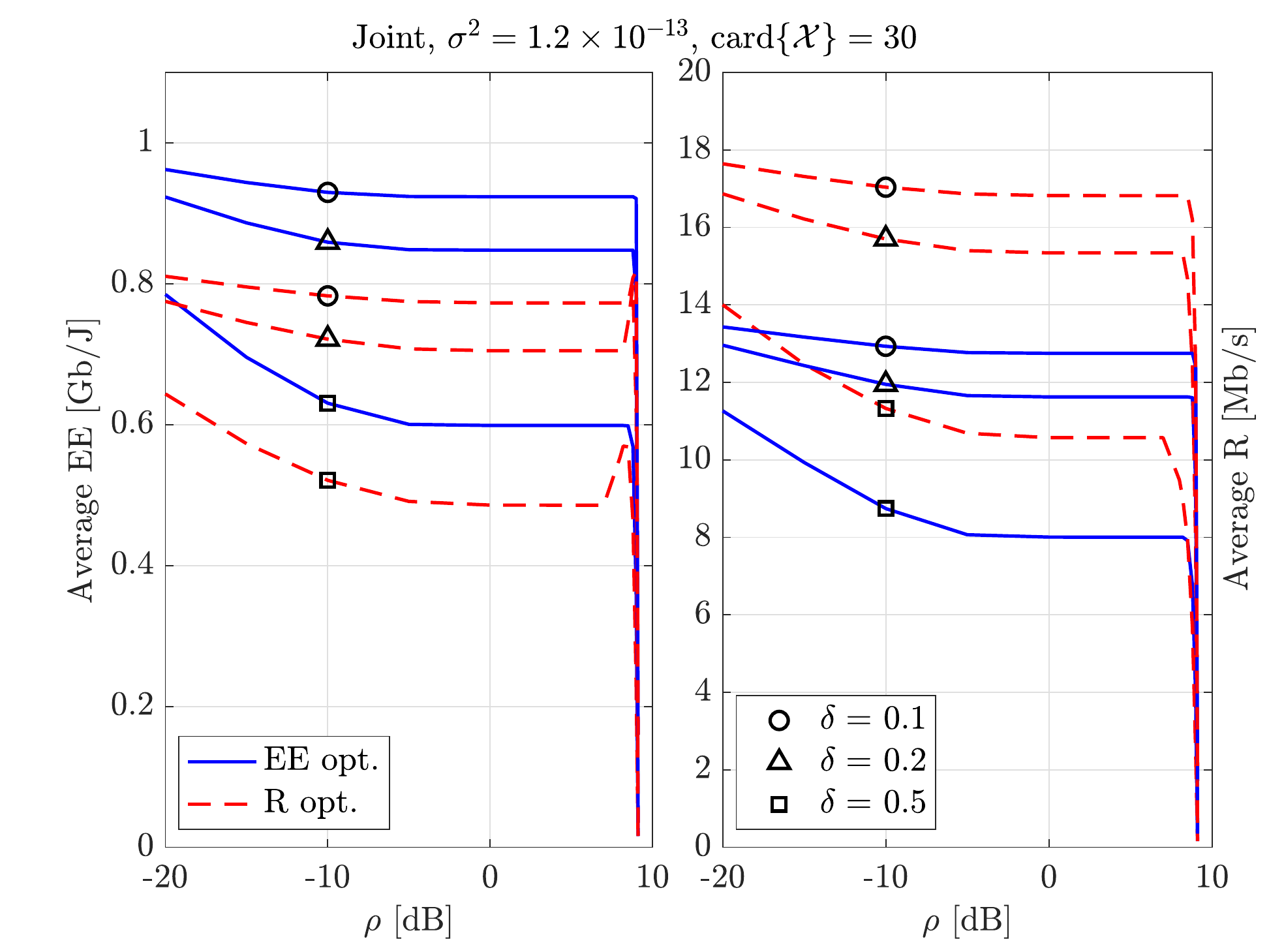}}
 \caption{Average energy efficiency (left plot) and average rate (right plot) versus the minimum required SDR for different design strategies (EE and rate optimization) and for different values of the density of the interference scatterers, when the intensity of the interference is $\sigma^2=1.2\times 10^{-13}$, and 30 resolution cells are protected.}\label{fig_07}
\end{figure}
\begin{figure}[t]
 \centering
 \hspace{-10pt}\centerline{\includegraphics[width=1.02\columnwidth]{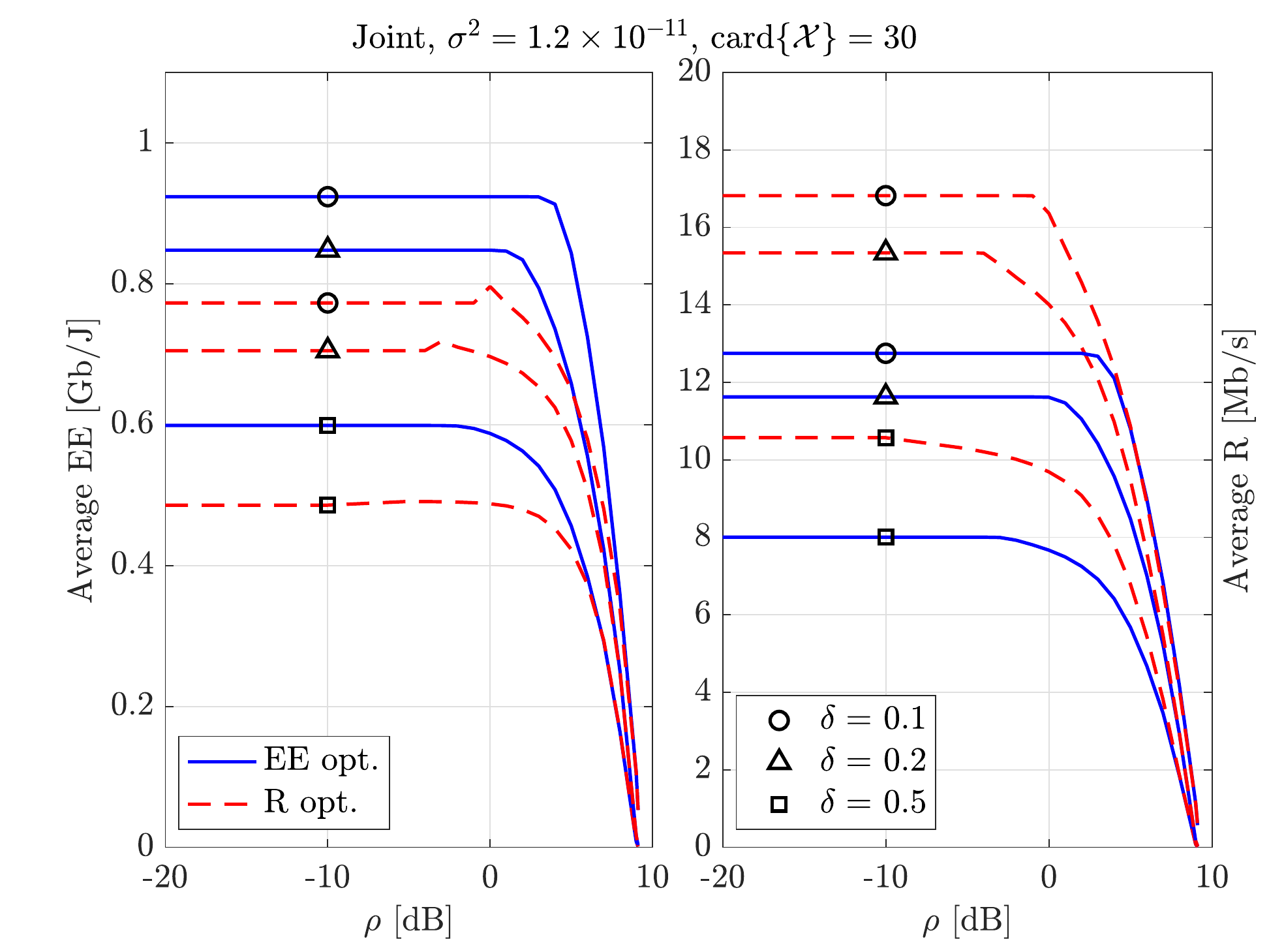}}
 \caption{Average energy efficiency (left plot) and average rate (right plot) versus the minimum required SDR for different design strategies (EE and rate optimization) and for different values of the density of the interference scatterers, when the intensity of the interference is $\sigma^2=1.2\times 10^{-11}$, and 30 resolution cells are protected.}\label{fig_08}
\end{figure}

Next we compare the proposed design strategy with that of optimizing the achievable communication rate $R$ in~\eqref{R_expr}, analyzed in~\cite{Grossi_2020_TSP}. In Figs.~\ref{fig_07} and~\ref{fig_08}, we report the performance in terms of average EE and average R versus $\rho$ for the two optimization strategies (referred to as \emph{EE opt.} and \emph{R opt.}) and for different values of of $\delta$, when $\sigma^2=1.2\times10^{-13}$ and $\sigma^2=1.2\times10^{-11}$, respectively. It can be seen that optimizing R causes a loss in terms of EE of about 20\% in a wide range of values of $\rho$; similarly, optimizing EE causes a rate loss of about 30\%. Therefore, the design strategy to be used depends on the performance metric that is most relevant in the application at hand.

\begin{figure}[t]
 \centering
 \centerline{\includegraphics[width=1.08\columnwidth]{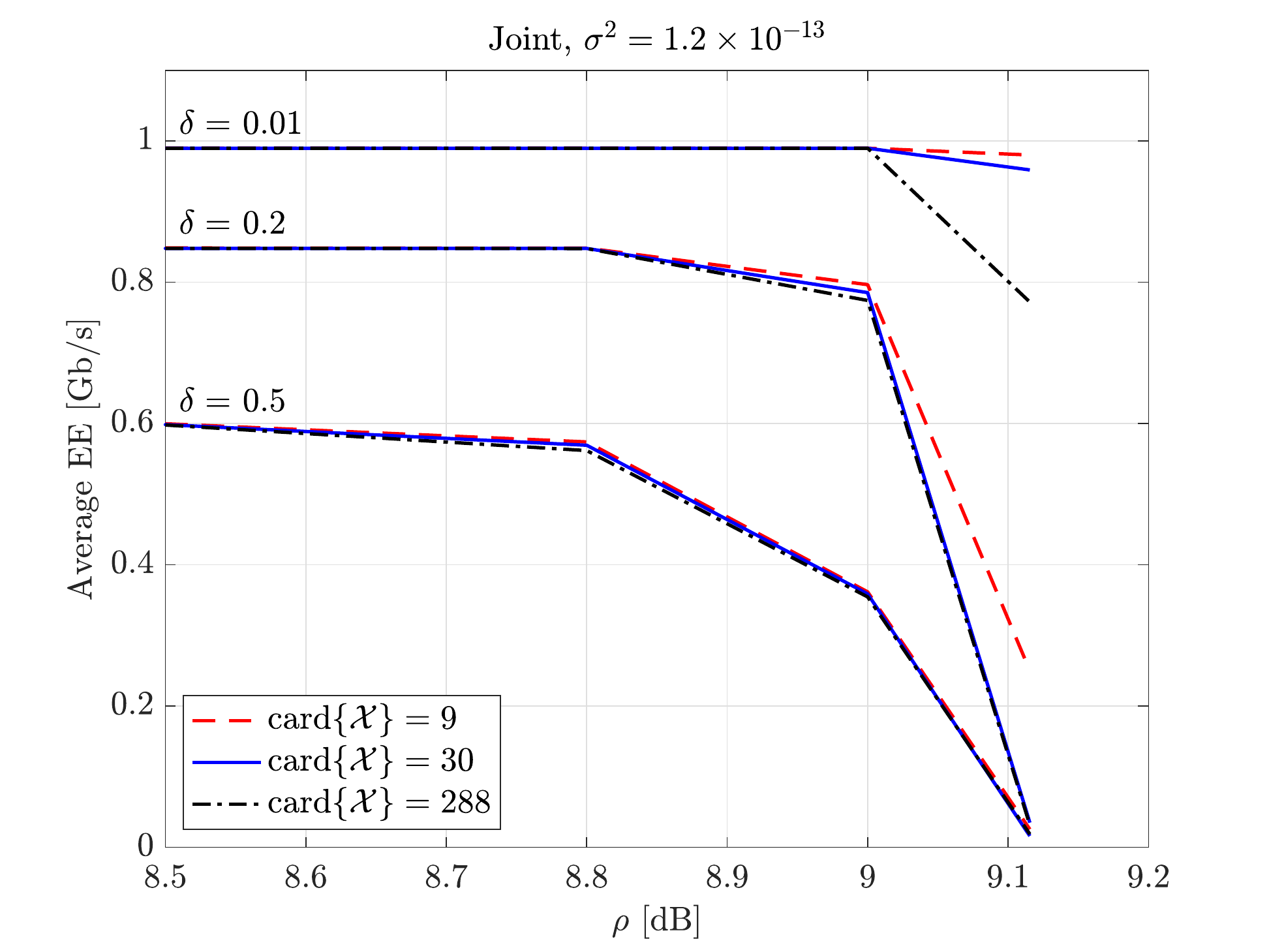}}
 \caption{Average energy efficiency versus the minimum required SDR in the joint design for different values of the number of protected resolution cells and of the density of the interference scatterers, $\delta$, when the intensity of the interference is $\sigma^2=1.2 \times 10^{-13}$.} \label{fig_09}
\end{figure}

\begin{figure}[t]
\centering
\centerline{\includegraphics[width=1.08\columnwidth]{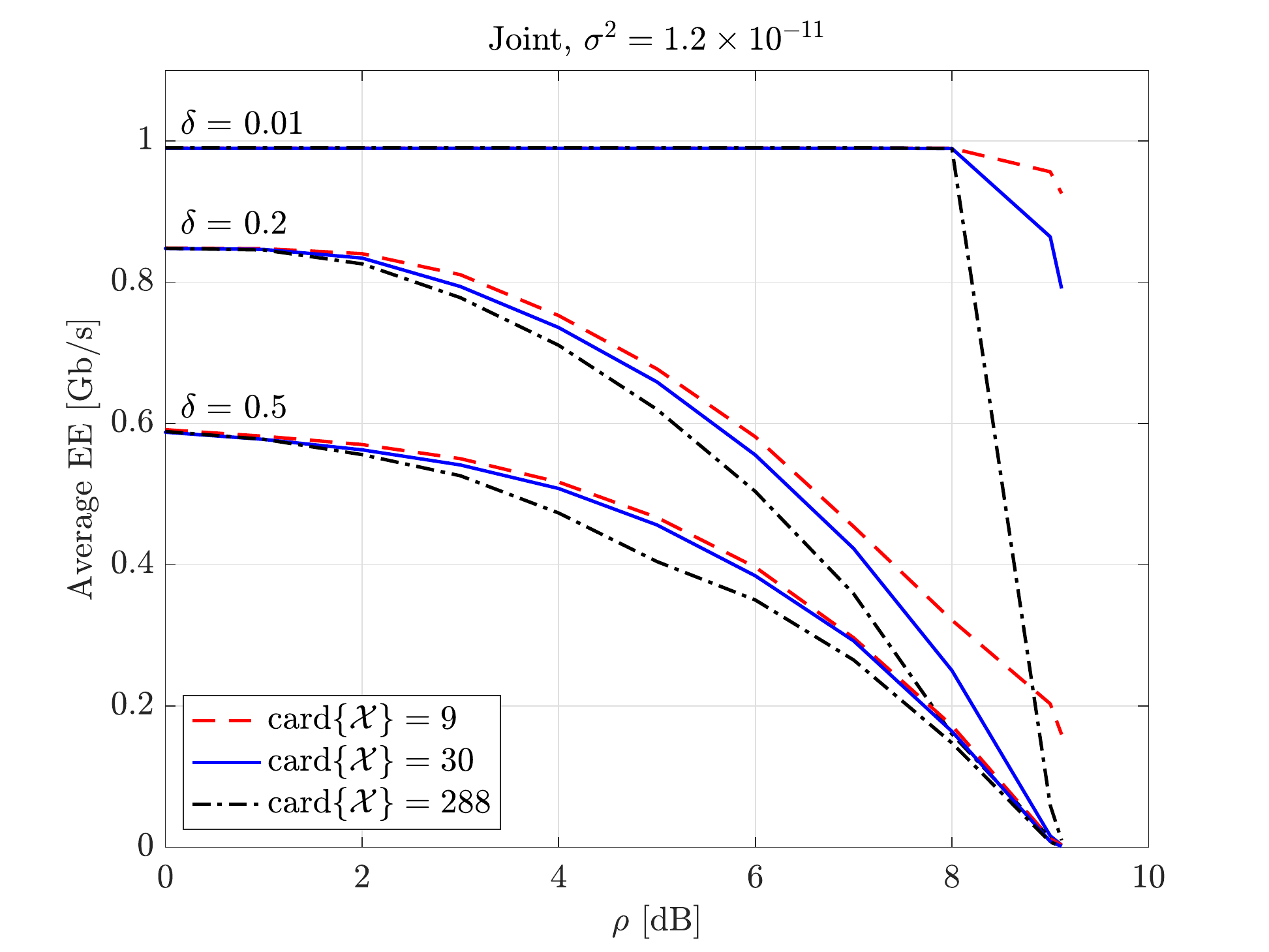}}
\caption{Average energy efficiency versus the minimum required SDR in the joint design for different values of the number of protected resolution cells and of the density of the interference scatterers, $\delta$, when the intensity of the interference is $\sigma^2=1.2 \times 10^{-11}$.} \label{fig_10}
\end{figure}

Finally, we show the impact of the number of protected radar cells on the system performance. In Figs.~\ref{fig_09} and~\ref{fig_10}, the average EE is plotted versus $\rho$ for different values of $\card \mathcal X$ and $\delta$, when $\sigma^2=1.2\times10^{-13}$ and $\sigma^2=1.2\times10^{-11}$, respectively. Clearly, the EE is decreasing with the cardinality of the set of protected cells, even if, in the inspected scenario, the difference in the performance is significant only when $\delta$ is very small and $\rho$ very large.

\section{Conclusion}\label{conclusion}

In this contribution, the problem of radar-communication spectrum sharing has been tackled in the context of green communications, i.e., assuming that the primary design goal is to maximize the energy efficiency while safeguarding the radar performance, fully leveraging the idea of cognition-based design for both active systems. The study relies on a quite realistic model, wherein a single radar range cell encompasses a whole cell of the wireless communication system: thus, the radar receiver is perturbated by the communication-induced interference, which may be of any kind ranging from a simple direct path to multi-path, on top of the signal-dependent reverberation produced by its own transmission. The communication receiver is conversely hit by the multiple rays of the rich scattering environment {\em locally} generated by the MIMO structure, along with the interference produced by random scatterers hit by the radar signal. 

The resulting constrained optimization problem is non-convex---due to both the chosen figure of merit and the signal-dependent clutter---and is approximately solved through the block coordinate ascent method, coupled with the Dinkelbach's algorithm and the quasi-Newton projected gradient method. A thorough performance analysis is also presented, aimed, on the one hand, at validating the iterative design procedure and assessing its convergence speed, and, on the other, at demonstrating the merits of the proposed approach. Interestingly, joint design allows a visible enlargement of the achievability region of communication EE and radar SDR pairs. Additionally, some light is shed on the strong interplay between these performance measures and such parameters as scatterers density and number of guaranteed range-azimuth cells of the radar system. 

Further possible developments include the adoption of different modulation formats (in particular, the proposed technique looks promising for optimizing the EE under multi-carrier modulations, which would allow to better exploit cognition in order to optimally allocate power in less interfered frequency bins), the inclusion of estimation errors, and the extension to the situation where multiple radars and/or multiple communication devices share the same spectrum. Also, many future works focused on specific application areas can be envisioned, with important practical and industrial outcomes.

\appendix

Let $\bm \mu = (\mu_1 \; \cdots\; \mu_U)\transp$ denote the Lagrangian multiplier associated with the linear matrix inequality constraints of Problem~\eqref{sub-prob_Dinkelbach}; then, the Lagrangian takes the form
 \begin{align}
 \mathcal L (\bm C, \bm \mu) &= \frac{\lambda}{\eta W \log_2\e}\trace \bm C - \ln \det \left( \bm I_{MN} + \bm F \bm C \right) \notag\\
 &\quad +\sum_{\ell=1}^U \mu_\ell \left(\trace( \bm E_\ell \bm C) - a_\ell \right)\notag\\
 &= \trace (\bm Z_{\bm \mu}^2 \bm C) - \ln \det \left( \bm I_{MN} + \bm F \bm C \right) - \bm a\transp \bm \mu \label{Lagrangian}
\end{align}
where $\bm a=(a_1 \;\cdots \; a_U)\transp$. The dual function is $g(\bm \mu)=\inf_{\bm C\succeq 0} \mathcal L (\bm C, \bm \mu)$ and can be found as follows.

Let $\bm X =\bm Z_{\bm \mu} \bm C \bm Z_{\bm \mu}$, then~\eqref{Lagrangian} can also be written as
 \begin{equation}
 \mathcal L (\bm C, \bm \mu) = \trace (\bm X) - \ln \det \left( \bm I_N + \bm Z_{\bm \mu}^{-1} \bm F\bm Z_{\bm \mu}^{-1} \bm X \right) - \bm a\transp \bm \mu  \label{Lagrangian_X}
\end{equation}
Since $\bm Z_{\bm \mu}$ and $\bm C$ are positive semidefinite, we have $\bm X\succeq0$, and the minimization of~\eqref{Lagrangian_X} over $\bm X\succeq0$ has a waterfilling-like solution. Specifically, since $\bm Z_{\bm \mu}^{-1} \bm F\bm Z_{\bm \mu}^{-1} =\bm V_{\bm \mu} \bm \Xi_{\bm \mu} \bm V_{\bm \mu}\herm$, by Hadamard's inequality~\cite[cfr. Th.~7.8.1]{Horn_Johnson_1985},~\eqref{Lagrangian_X} is minimized if $\Delta$ eigenvectors of $\bm X$ are matched to $\bm V_{\bm \mu}$, and the corresponding eigenvalues, say $\{x_\ell\}_{\ell=1}^\Delta$, are optimized; the remaining $MN-\Delta$ eigenvalues, should instead be set equal to zero, as they can only increase the trace term in~\eqref{Lagrangian_X}. The problem is, therefore,
\begin{equation}
 \max_{\substack{\{x_\ell\}_{\ell=1}^\Delta: \\ x_\ell\geq0,\forall \ell}}  \sum_{\ell=1}^\Delta \bigl( x_\ell - \ln ( 1+\xi_{\bm \mu, \ell} x_\ell ) \bigr)
\end{equation}
and the solution is simply $x_\ell=( 1- \xi_{\bm \mu, \ell}^{-1})^+$, $\ell=1,\ldots,\Delta$, so that $\bm X= \bm V_{\bm \mu} ( \bm I_\Delta - \bm \Xi_{\bm \mu}^{-1})^+ \bm V_{\bm \mu}\herm$. In this case, $\bm C=\bm Z_{\bm \mu}^{-1} \bm X \bm Z_{\bm \mu}^{-1}$ as in~\eqref{sol_C}, and $g(\bm \mu)$ takes the form in~\eqref{dual_fun}.

Since Problem~\eqref{sub-prob_C}, and therefore Problem~\eqref{sub-prob_Dinkelbach}, is strictly feasible, strong duality holds~\cite{Boyd_Vandenberghe_2004}: there exist $\bm \mu$ that is optimal for the dual problem in~\eqref{dual_prob} with dual objective equal to the optimal value of the primal problem in~\eqref{sub-prob_Dinkelbach}; moreover, the solution of Problem~\eqref{sub-prob_Dinkelbach} can be recovered from the dual optimal variables using~\eqref{sol_C}.

\bibliographystyle{IEEEtran}
\bibliography{IEEEabrv,references}

\begin{thebibliography}{10}
\providecommand{\url}[1]{#1}
\csname url@samestyle\endcsname
\providecommand{\newblock}{\relax}
\providecommand{\bibinfo}[2]{#2}
\providecommand{\BIBentrySTDinterwordspacing}{\spaceskip=0pt\relax}
\providecommand{\BIBentryALTinterwordstretchfactor}{4}
\providecommand{\BIBentryALTinterwordspacing}{\spaceskip=\fontdimen2\font plus
\BIBentryALTinterwordstretchfactor\fontdimen3\font minus
  \fontdimen4\font\relax}
\providecommand{\BIBforeignlanguage}[2]{{%
\expandafter\ifx\csname l@#1\endcsname\relax
\typeout{** WARNING: IEEEtran.bst: No hyphenation pattern has been}%
\typeout{** loaded for the language `#1'. Using the pattern for}%
\typeout{** the default language instead.}%
\else
\language=\csname l@#1\endcsname
\fi
#2}}
\providecommand{\BIBdecl}{\relax}
\BIBdecl
\renewcommand{\BIBentryALTinterwordstretchfactor}{4}

\bibitem{Lambert_2012}
S.~Lambert, W.~{Van Heddeghem}, W.~Vereecken, B.~Lannoo, D.~Colle, and
  M.~Pickavet, ``Worldwide electricity consumption of communication networks,''
  \emph{Opt. Express}, vol.~20, no.~26, pp. B513--B524, Dec. 2012.

\bibitem{5g-ppp}
{The 5G Infrastructure Public Private Partnership (5G PPP)}, ``5{G} key
  performance indicators,'' https://5g-ppp.eu/kpis/, Accessed: April 1, 2020.

\bibitem{ITU_2017}
ITU-R, ``Minimum requirements related to technical performance for {IMT}-2020
  radio interface(s),'' International Telecommunication Unit ({ITU}), Tech.
  Rep. M.2410-0, Nov. 2017.

\bibitem{Kwon_1986}
H.~Kwon and T.~Birdsall, ``Channel capacity in bits per joule,'' \emph{{IEEE}
  J. Ocean. Eng.}, vol.~11, no.~1, pp. 97--99, Jan. 1986.

\bibitem{Isheden_2012}
C.~{Isheden}, Z.~{Chong}, E.~{Jorswieck}, and G.~{Fettweis}, ``Framework for
  link-level energy efficiency optimization with informed transmitter,''
  \emph{{IEEE} Trans. Wireless Commun.}, vol.~11, no.~8, pp. 2946--2957, 2012.

\bibitem{Mammela_2017}
A.~Mammela and A.~Anttonen, ``Why will computing power need particular
  attention in future wireless devices?'' \emph{{IEEE} Circuits Syst. Mag.},
  vol.~17, no.~1, pp. 12--26, 2017.

\bibitem{Correia_2010}
L.~M. {Correia}, D.~{Zeller}, O.~{Blume}, D.~{Ferling}, Y.~{Jading},
  I.~{Gódor}, G.~{Auer}, and L.~V. {Der Perre}, ``Challenges and enabling
  technologies for energy aware mobile radio networks,'' \emph{{IEEE} Commun.
  Mag.}, vol.~48, no.~11, pp. 66--72, 2010.

\bibitem{Chen_2011}
Y.~{Chen}, S.~{Zhang}, S.~{Xu}, and G.~Y. {Li}, ``Fundamental trade-offs on
  green wireless networks,'' \emph{{IEEE} Commun. Mag.}, vol.~49, no.~6, pp.
  30--37, 2011.

\bibitem{Vereecken_2011}
W.~Vereecken, W.~V. Heddeghem, M.~Deruyck, B.~Puype, B.~Lannoo, W.~Joseph,
  D.~Colle, L.~Martens, and P.~Demeester, ``Power consumption in
  telecommunication networks: overview and reduction strategies,'' \emph{{IEEE}
  Commun. Mag.}, vol.~49, no.~6, pp. 62--69, Jun. 2011.

\bibitem{Hinton_2011}
K.~Hinton, J.~Baliga, M.~Feng, R.~Ayre, and R.~S. Tucker, ``Power consumption
  and energy efficiency in the internet,'' \emph{{IEEE} Netw.}, vol.~25, no.~2,
  pp. 6--12, Mar./Apr. 2011.

\bibitem{Han_2011}
C.~Han, T.~Harrold, S.~Armour, I.~Krikidis, S.~Videv, P.~M. Grant, H.~Haas,
  J.~S. Thompson, I.~Ku, C.~Wang, T.~A. Le, M.~R. Nakhai, J.~Zhang, and
  L.~Hanzo, ``Green radio: radio techniques to enable energy-efficient wireless
  networks,'' \emph{{IEEE} Commun. Mag.}, vol.~49, no.~6, pp. 46--54, Jun.
  2011.

\bibitem{Bianzino_2012}
A.~P. Bianzino, C.~Chaudet, D.~Rossi, and J.~Rougier, ``A survey of green
  networking research,'' \emph{{IEEE} Commun. Surveys Tuts.}, vol.~14, no.~1,
  pp. 3--20, 2012.

\bibitem{Feng_2013}
D.~Feng, C.~Jiang, G.~Lim, L.~J. Cimini, G.~Feng, and G.~Y. Li, ``A survey of
  energy-efficient wireless communications,'' \emph{{IEEE} Commun. Surveys
  Tuts.}, vol.~15, no.~1, pp. 167--178, 2013.

\bibitem{Miao_2012}
G.~{Miao}, N.~{Himayat}, G.~Y. {Li}, and S.~{Talwar}, ``Low-complexity
  energy-efficient scheduling for uplink {OFDMA},'' \emph{{IEEE} Trans.
  Commun.}, vol.~60, no.~1, pp. 112--120, 2012.

\bibitem{Xiong_2011}
C.~{Xiong}, G.~Y. {Li}, S.~{Zhang}, Y.~{Chen}, and S.~{Xu}, ``Energy- and
  spectral-efficiency tradeoff in downlink {OFDMA} networks,'' \emph{{IEEE}
  Trans. Wireless Commun.}, vol.~10, no.~11, pp. 3874--3886, 2011.

\bibitem{Ng_2013}
D.~W.~K. {Ng}, E.~S. {Lo}, and R.~{Schober}, ``Energy-efficient resource
  allocation in {OFDMA} systems with hybrid energy harvesting base station,''
  \emph{{IEEE} Trans. Wireless Commun.}, vol.~12, no.~7, pp. 3412--3427, 2013.

\bibitem{Fang_2016}
F.~{Fang}, H.~{Zhang}, J.~{Cheng}, and V.~C.~M. {Leung}, ``Energy-efficient
  resource allocation for downlink non-orthogonal multiple access network,''
  \emph{{IEEE} Trans. Commun.}, vol.~64, no.~9, pp. 3722--3732, 2016.

\bibitem{Soh_2013}
Y.~S. {Soh}, T.~Q.~S. {Quek}, M.~{Kountouris}, and H.~{Shin}, ``Energy
  efficient heterogeneous cellular networks,'' \emph{{IEEE} J. Sel. Areas
  Commun.}, vol.~31, no.~5, pp. 840--850, May 2013.

\bibitem{Xu_2013}
J.~{Xu} and L.~{Qiu}, ``Energy efficiency optimization for {MIMO} broadcast
  channels,'' \emph{{IEEE} Trans. Wireless Commun.}, vol.~12, no.~2, pp.
  690--701, 2013.

\bibitem{Nguyen_2013}
D.~{Nguyen}, L.~{Tran}, P.~{Pirinen}, and M.~{Latva-aho}, ``Precoding for full
  duplex multiuser mimo systems: Spectral and energy efficiency maximization,''
  \emph{{IEEE} Trans. Signal Process.}, vol.~61, no.~16, pp. 4038--4050, 2013.

\bibitem{Ngo_2013}
H.~Q. {Ngo}, E.~G. {Larsson}, and T.~L. {Marzetta}, ``Energy and spectral
  efficiency of very large multiuser mimo systems,'' \emph{{IEEE} Trans.
  Commun.}, vol.~61, no.~4, pp. 1436--1449, 2013.

\bibitem{Bjornson_2014}
E.~{Bj\:ornson}, J.~{Hoydis}, M.~{Kountouris}, and M.~{Debbah}, ``Massive
  {MIMO} systems with non-ideal hardware: Energy efficiency, estimation, and
  capacity limits,'' \emph{{IEEE} Trans. Inf. Theory}, vol.~60, no.~11, pp.
  7112--7139, Nov. 2014.

\bibitem{Huang_2019}
C.~{Huang}, A.~{Zappone}, G.~C. {Alexandropoulos}, M.~{Debbah}, and C.~{Yuen},
  ``Reconfigurable intelligent surfaces for energy efficiency in wireless
  communication,'' \emph{{IEEE} Trans. Wireless Commun.}, vol.~18, no.~8, pp.
  4157--4170, 2019.

\bibitem{Ren_2016}
J.~{Ren}, Y.~{Zhang}, N.~{Zhang}, D.~{Zhang}, and X.~{Shen}, ``Dynamic channel
  access to improve energy efficiency in cognitive radio sensor networks,''
  \emph{{IEEE} Trans. Wireless Commun.}, vol.~15, no.~5, pp. 3143--3156, 2016.

\bibitem{Fodor_2012}
G.~{Fodor}, E.~{Dahlman}, G.~{Mildh}, S.~{Parkvall}, N.~{Reider}, G.~{Miklós},
  and Z.~{Turányi}, ``Design aspects of network assisted device-to-device
  communications,'' \emph{{IEEE} Commun. Mag.}, vol.~50, no.~3, pp. 170--177,
  2012.

\bibitem{Feng_2014}
D.~{Feng}, L.~{Lu}, Y.~{Yuan-Wu}, G.~Y. {Li}, S.~{Li}, and G.~{Feng},
  ``Device-to-device communications in cellular networks,'' \emph{{IEEE}
  Commun. Mag.}, vol.~52, no.~4, pp. 49--55, 2014.

\bibitem{Wang_2016}
K.~{Wang}, Y.~{Wang}, Y.~{Sun}, S.~{Guo}, and J.~{Wu}, ``Green industrial
  internet of things architecture: An energy-efficient perspective,''
  \emph{{IEEE} Commun. Mag.}, vol.~54, no.~12, pp. 48--54, 2016.

\bibitem{Zhang_2016}
D.~{Zhang}, Z.~{Zhou}, S.~{Mumtaz}, J.~{Rodriguez}, and T.~{Sato}, ``One
  integrated energy efficiency proposal for 5g iot communications,''
  \emph{{IEEE} Internet Things J.}, vol.~3, no.~6, pp. 1346--1354, 2016.

\bibitem{Buzzi_2016}
S.~{Buzzi}, C.-L. {I}, T.~E. {Klein}, H.~V. {Poor}, C.~{Yang}, and
  A.~{Zappone}, ``A survey of energy-efficient techniques for 5g networks and
  challenges ahead,'' \emph{{IEEE} J. Sel. Areas Commun.}, vol.~34, no.~4, pp.
  697--709, Apr. 2016.

\bibitem{Zappone_2016}
A.~{Zappone}, L.~{Sanguinetti}, G.~{Bacci}, E.~{Jorswieck}, and M.~{Debbah},
  ``Energy-efficient power control: A look at 5g wireless technologies,''
  \emph{{IEEE} Trans. Signal Process.}, vol.~64, no.~7, pp. 1668--1683, Apr.1
  2016.

\bibitem{QUALCOMM_2013}
{Qualcomm Incorporated}, ``$1000\times$ mobile data challenge,''
  https://www.qualcomm.com/media/documents/files/1000x-mobile-data-challenge.pdf,
  2013.

\bibitem{CISCO_internet_2020}
{Cisco Systems Inc.}, ``Cisco annual internet report (2018--2023),''
  https://www.cisco.com/c/en/us/solutions/collateral/executive-perspectives/annual-internet-report/white-paper-c11-741490.pdf,
  2020.

\bibitem{Griffiths_2015}
H.~Griffiths, L.~Cohen, S.~Watts, E.~Mokole, C.~Baker, M.~Wicks, and S.~D.
  Blunt, ``Radar spectrum engineering and management: Technical and regulatory
  issues,'' \emph{Proc. {IEEE}}, vol. 103, no.~1, pp. 85--102, 2015.

\bibitem{Liu_2020}
F.~{Liu}, C.~{Masouros}, A.~{Petropulu}, H.~{Griffiths}, and L.~{Hanzo},
  ``Joint radar and communication design: Applications, state-of-the-art, and
  the road ahead,'' \emph{{IEEE} Trans. Commun.}, to appear.

\bibitem{SSPARC_2013}
{Defense Advanced Research Projects Agency (DARPA)}, ``Shared spectrum access
  for radar and communications ({SSPARC}),'' {DARPA} {BAA}-13-24,
  www.darpa.mil, 2013.

\bibitem{Jacyna_2016}
G.~M. Jacyna, B.~Fell, and D.~McLemore, ``A high-level overview of fundamental
  limits studies for the {DARPA} {SSPARC} program,'' in \emph{{IEEE} Radar
  Conf. ({RadarConf})}, Philadelphia, {PA}, {USA}, May 2016, pp. 198--203.

\bibitem{Chiryath_2016}
A.~R. Chiriyath, B.~Paul, G.~M. Jacyna, and D.~W. Bliss, ``Inner bounds on
  performance of radar and communications co-existence,'' \emph{{IEEE} Trans.
  Signal Process.}, vol.~64, no.~2, pp. 464--474, Jan.15 2016.

\bibitem{Liu_2018}
F.~{Liu}, L.~{Zhou}, C.~{Masouros}, A.~{Li}, W.~{Luo}, and A.~{Petropulu},
  ``Toward dual-functional radar-communication systems: Optimal waveform
  design,'' \emph{{IEEE} Trans. Signal Process.}, vol.~66, no.~16, pp.
  4264--4279, Aug.15 2018.

\bibitem{Mealey_1963}
R.~M. {Mealey}, ``A method for calculating error probabilities in a radar
  communication system,'' \emph{IEEE Trans. Space Electron. and Telemetry},
  vol.~9, no.~2, pp. 37--42, Jun. 1963.

\bibitem{Blunt_2010}
S.~D. Blunt, P.~Yathan, and J.~Stiles, ``Intrapulse radar-embedded
  communications,'' \emph{{IEEE} Trans. Aerosp. Electron. Syst.}, vol.~46,
  no.~3, pp. 1185--1200, 2010.

\bibitem{Hassanien_2016}
A.~Hassanien, M.~Amin, Y.~Zhang, and F.~Ahmad, ``Dual-function
  radar-communications: Information embedding using sidelobe control and
  waveform diversity,'' \emph{{IEEE} Trans. Signal Process.}, vol.~64, no.~8,
  pp. 2168--2181, 2016.

\bibitem{Wang_Hassanien_2019}
X.~Wang, A.~Hassanien, and M.~G. Amin, ``Dual-function {MIMO} radar
  communications system design via sparse array optimization,'' \emph{{IEEE}
  Trans. Aerosp. Electron. Syst.}, vol.~55, no.~3, pp. 1213--1226, Jun. 2019.

\bibitem{Sturm_2011}
C.~{Sturm} and W.~{Wiesbeck}, ``Waveform design and signal processing aspects
  for fusion of wireless communications and radar sensing,'' \emph{Proc.
  {IEEE}}, vol.~99, no.~7, pp. 1236--1259, Jul. 2011.

\bibitem{Zhou_2019}
S.~Zhou, X.~Liang, Y.~Yu, and H.~Liu, ``Joint radar-communications co-use
  waveform design using optimized phase perturbation,'' \emph{{IEEE} Trans.
  Aerosp. Electron. Syst.}, vol.~55, no.~3, pp. 1227--1240, Jun. 2019.

\bibitem{Grossi_2017_radarcon}
E.~Grossi, M.~Lops, L.~Venturino, and A.~Zappone, ``Opportunistic automotive
  radar using the {IEEE} 802.11ad standard,'' in \emph{{IEEE} Radar Conf.
  (Radar{C}onf)}, Seattle, {WA}, {USA}, May 2017, pp. 1196--1200.

\bibitem{Grossi_2017_VTC}
E.~Grossi, M.~Lops, L.~Venturino, and A.~Zappone, ``Opportunistic radar in
  {IEEE} 802.11ad vehicular networks,'' in \emph{Veh. Technol. Conf. ({VTC}
  Spring)}, Sydney, {NSW}, Australia, Jun. 2017.

\bibitem{Grossi_2018_TSP}
E.~Grossi, M.~Lops, L.~Venturino, and A.~Zappone, ``Opportunistic radar in
  802.11ad networks,'' \emph{{IEEE} Trans. Signal Process.}, vol.~66, no.~9,
  pp. 2441--2454, 2018.

\bibitem{Kumari_2018}
P.~{Kumari}, J.~{Choi}, N.~{Gonz\'alez-Prelcic}, and R.~W. {Heath}, ``{IEEE}
  802.11ad-based radar: An approach to joint vehicular communication-radar
  system,'' \emph{{IEEE} Trans. Veh. Technol.}, vol.~67, no.~4, pp. 3012--3027,
  Apr. 2018.

\bibitem{Grossi_2019_asilomar}
E.~Grossi, M.~Lops, and L.~Venturino, ``Detection and localization of multiple
  targets in ieee 802.11ad networks,'' in \emph{Asilomar Conference on Signals,
  Systems, and Computers}, Pacific Grove, {CA}, {USA}, Oct. 2019, pp. 947--951.

\bibitem{Grossi_2019_camsap_b}
E.~Grossi, M.~Lops, and L.~Venturino, ``An iterative interference cancellation
  algorithm for opportunistic sensing in ieee 802.11ad networks,'' in
  \emph{{IEEE} Int. Workshop Comput. Adv. Multi-Sensor Adapt. Process.
  (CAMSAP)}, Le Gosier, Guadeloupe, Dec. 2019, pp. 141--145.

\bibitem{Grossi_2020_TWC}
E.~{Grossi}, M.~{Lops}, and L.~{Venturino}, ``Adaptive detection and
  localization exploiting the {IEEE} 802.11ad standard,'' \emph{{IEEE} Trans.
  Wireless Commun.}, vol.~19, no.~7, pp. 4394--4407, 2020.

\bibitem{Wang_2008}
L.~S. Wang, J.~P. Mcgeehan, C.~Williams, and A.~Doufexi, ``Application of
  cooperative sensing in radar-communications coexistence,'' \emph{{IET}
  Commun.}, vol.~2, no.~6, pp. 856--868, Jul. 2008.

\bibitem{Saruthirathanaworakun_2012}
R.~{Saruthirathanaworakun}, J.~M. {Peha}, and L.~M. {Correia}, ``Opportunistic
  sharing between rotating radar and cellular,'' \emph{{IEEE} J. Sel. Areas
  Commun.}, vol.~30, no.~10, pp. 1900--1910, Nov. 2012.

\bibitem{Hessar_2016}
F.~Hessar and S.~Roy, ``Spectrum sharing between a surveillance radar and
  secondary {W}i-{F}i networks,'' \emph{{IEEE} Trans. Aerosp. Electron. Syst.},
  vol.~52, no.~3, pp. 1434--1448, Jun. 2016.

\bibitem{Raymond_2016}
S.~Raymond, A.~Abubakari, and H.~Jo, ``Coexistence of power-controlled cellular
  networks with rotating radar,'' \emph{{IEEE} J. Sel. Areas Commun.}, vol.~34,
  no.~10, pp. 2605--2616, Oct. 2016.

\bibitem{Li_2017}
B.~Li and A.~P. Petropulu, ``Joint transmit designs for coexistence of {MIMO}
  wireless communications and sparse sensing radars in clutter,'' \emph{{IEEE}
  Trans. Aerosp. Electron. Syst.}, vol.~53, no.~6, pp. 2846--2864, 2017.

\bibitem{Liu_2019}
F.~{Liu}, A.~{Garcia-Rodriguez}, C.~{Masouros}, and G.~{Geraci}, ``Interfering
  channel estimation in radar-cellular coexistence: How much information do we
  need?'' \emph{{IEEE} Trans. Wireless Commun.}, vol.~18, no.~9, pp.
  4238--4253, Sep. 2019.

\bibitem{Deng_2013}
H.~Deng and B.~Himed, ``Interference mitigation processing for spectrum-sharing
  between radar and wireless communications systems,'' \emph{{IEEE} Trans.
  Aerosp. Electron. Syst.}, vol.~49, no.~3, pp. 1911--1919, Jul. 2013.

\bibitem{Geng_2015}
Z.~Geng, H.~Deng, and B.~Himed, ``Adaptive radar beamforming for interference
  mitigation in radar-wireless spectrum sharing,'' \emph{{IEEE} Signal Process.
  Lett.}, vol.~22, no.~4, pp. 484--488, Apr. 2015.

\bibitem{Mahal_2017}
J.~A. Mahal, A.~Khawar, A.~Abdelhadi, and T.~C. Clancy, ``Spectral coexistence
  of {MIMO} radar and {MIMO} cellular system,'' \emph{{IEEE} Trans. Aerosp.
  Electron. Syst.}, vol.~53, no.~2, pp. 655--668, Apr. 2017.

\bibitem{Cheng_2018}
Z.~Cheng, C.~Han, B.~Liao, Z.~He, and J.~Li, ``Communication-aware waveform
  design for {MIMO} radar with good transmit beampattern,'' \emph{{IEEE} Trans.
  Signal Process.}, vol.~66, no.~21, pp. 5549--5562, Nov. 2018.

\bibitem{Shi_2018}
C.~Shi, F.~Wang, M.~Sellathurai, J.~Zhou, and S.~Salous, ``Power
  minimization-based robust {OFDM} radar waveform design for radar and
  communication systems in coexistence,'' \emph{{IEEE} Trans. Signal Process.},
  vol.~66, no.~5, pp. 1316--1330, Mar. 2018.

\bibitem{Bica_2019}
M.~Bic\v{a} and V.~Koivunen, ``Radar waveform optimization for target parameter
  estimation in cooperative radar-communications systems,'' \emph{{IEEE} Trans.
  Aerosp. Electron. Syst.}, vol.~55, no.~5, pp. 2314--2326, Oct. 2019.

\bibitem{Nartasilpa_2018}
N.~Nartasilpa, A.~Salim, D.~Tuninetti, and N.~Devroye, ``Communications system
  performance and design in the presence of radar interference,'' \emph{{IEEE}
  Trans. Commun.}, vol.~66, no.~9, pp. 4170--4185, Sep. 2018.

\bibitem{Li_2016}
B.~Li, A.~P. Petropulu, and W.~Trappe, ``Optimum co-design for spectrum sharing
  between matrix completion based {MIMO} radars and a {MIMO} communication
  system,'' \emph{{IEEE} Trans. Signal Process.}, vol.~64, no.~17, pp.
  4562--4575, Sep. 2016.

\bibitem{Zheng_2017}
L.~Zheng, M.~Lops, X.~Wang, E.~Grossi, and J.~Qian, ``Joint design of
  co-existing communication system and pulsed radar,'' in \emph{{IEEE} Int.
  Workshop Comput. Adv. Multi-Sensor Adapt. Process. (CAMSAP)}, Curacao,
  Netherlands Antilles, Dec. 2017, pp. 1--5.

\bibitem{Zheng_2018_a}
L.~Zheng, M.~Lops, X.~Wang, and E.~Grossi, ``Joint design of overlaid
  communication systems and pulsed radars,'' \emph{{IEEE} Trans. Signal
  Process.}, vol.~66, no.~1, pp. 139--154, Jan. 2018.

\bibitem{Grossi_2018}
E.~Grossi, M.~Lops, and L.~Venturino, ``Joint design of communication and radar
  transceiver in spectrum-sharing architectures,'' in \emph{Asilomar Conference
  on Signals, Systems, and Computers}, Pacific Grove, {CA}, {USA}, Oct. 2018,
  pp. 962--966.

\bibitem{Rihan_2018}
M.~Rihan and L.~Huang, ``Optimum co-design of spectrum sharing between {MIMO}
  radar and {MIMO} communication systems: An interference alignment approach,''
  \emph{{IEEE} Trans. Veh. Technol.}, vol.~7, no.~12, pp. 11\,667--11\,680,
  Dec. 2018.

\bibitem{Grossi_2019_camsap_a}
E.~Grossi, M.~Lops, and L.~Venturino, ``Spectrum-sharing between a surveillance
  radar and a mimo communication system in cluttered environments,'' in
  \emph{{IEEE} Int. Workshop Comput. Adv. Multi-Sensor Adapt. Process.
  (CAMSAP)}, Le Gosier, Guadeloupe, Dec. 2019, pp. 371--375.

\bibitem{Wang_2019}
F.~Wang, H.~Li, and M.~A. Govoni, ``Power allocation and co-design of
  multicarrier communication and radar systems for spectral coexistence,''
  \emph{{IEEE} Trans. Signal Process.}, vol.~67, no.~14, pp. 3818--3831, Jul.
  2019.

\bibitem{Cheng_2019}
Z.~{Cheng}, B.~{Liao}, S.~{Shi}, Z.~{He}, and J.~{Li}, ``Co-design for overlaid
  mimo radar and downlink miso communication systems via cram\'er–rao bound
  minimization,'' \emph{{IEEE} Trans. Signal Process.}, vol.~67, no.~24, pp.
  6227--6240, Dec.15 2019.

\bibitem{Grossi_2020_TSP}
E.~{Grossi}, M.~{Lops}, and L.~{Venturino}, ``Joint design of surveillance
  radar and mimo communication in cluttered environments,'' \emph{{IEEE} Trans.
  Signal Process.}, vol.~68, no.~1, pp. 1544--1557, Dec. 2020.

\bibitem{Grossi_2020_SAM}
E.~Grossi, M.~Lops, and L.~Venturino, ``Energy efficient communication with
  radar spectrum sharing,'' in \emph{{IEEE} Sensor Array Multichannel Signal
  Process. Workshop ({SAM})}, Hangzhou, China, Jun. 2020.

\bibitem{Skolnik_2001}
M.~I. Skolnik, \emph{Introduction to Radar Systems}.\hskip 1em plus 0.5em minus
  0.4em\relax McGraw-Hill, 2001.

\bibitem{Rudge_1983}
D.~E.~N. Davies, K.~Corless, D.~S. Hicks, and K.~Milne, ``Array signal
  processing,'' in \emph{The Handbook of Antenna Design, Vol.~2}, A.~Rudge,
  K.~Milne, A.~Olver, and P.~Knight, Eds.\hskip 1em plus 0.5em minus
  0.4em\relax London, UK: Peter Peregrinus Ltd, 1983.

\bibitem{Wirth_2013}
W.-D. Wirth, \emph{Radar Techniques Using Array Antennas}, 2nd~ed.\hskip 1em
  plus 0.5em minus 0.4em\relax IEE, 2013.

\bibitem{Skolnik_2002_a}
M.~Skolnik, ``Role of radar in microwaves,'' \emph{{IEEE} Trans. Microw. Theory
  Techn.}, vol.~50, no.~3, pp. 625--632, Mar. 2002.

\bibitem{Skolnik_2008}
M.~I. Skolnik, \emph{Radar Handbook}.\hskip 1em plus 0.5em minus 0.4em\relax
  McGraw-Hill, 2008.

\bibitem{Deng_2004}
H.~Deng, ``Polyphase code design for orthogonal netted radar systems,''
  \emph{{IEEE} Trans. Signal Process.}, vol.~52, no.~11, pp. 3126--3135, Nov.
  2004.

\bibitem{Zappone_2015}
A.~{Zappone} and E.~{Jorswieck}, \emph{Energy Efficiency in Wireless Networks
  via Fractional Programming Theory}.\hskip 1em plus 0.5em minus 0.4em\relax
  Now Foundations and Trends, 2015.

\bibitem{Bertsekas_1999}
D.~P. Bertsekas, \emph{Nonlinear Programming}.\hskip 1em plus 0.5em minus
  0.4em\relax Athena Scientific, 1999.

\bibitem{Jagannathan_1966}
R.~Jagannathan, ``On some properties of programming problems in parametric form
  pertaining to fractional programming,'' \emph{Manage. Sci.}, vol.~12, no.~7,
  pp. 609--615, Mar. 1966.

\bibitem{Dinkelbach_1967}
W.~Dinkelbach, ``On nonlinear fractional programming,'' \emph{Manage. Sci.},
  vol.~13, no.~7, pp. 492--498, Jul. 1967.

\bibitem{Crouzeix_1991}
J.~P. Crouzeix and J.~A. Ferland, ``Algorithms for generalized fractional
  programming,'' \emph{Math. Program.}, vol.~52, pp. 191--207, 1991.

\bibitem{Nocedal_2006}
J.~Nocedal and S.~Wright, \emph{Numerical Optimization}.\hskip 1em plus 0.5em
  minus 0.4em\relax Springer-Verlag, 2006.

\bibitem{Schaible_1976}
S.~Schaible, ``Fractional programming. {II}, on dinkelbach's algorithm,''
  \emph{Manage. Sci.}, vol.~22, no.~8, pp. 868--873, Apr. 1976.

\bibitem{Boyd_Vandenberghe_2004}
S.~Boyd and L.~Vandenberghe, \emph{Convex optimization}.\hskip 1em plus 0.5em
  minus 0.4em\relax Cambridge University Press, 2004.

\bibitem{Zhang_2010}
R.~Zhang, Y.~Liang, and S.~Cui, ``Dynamic resource allocation in cognitive
  radio networks,'' \emph{{IEEE} Signal Process. Mag.}, vol.~27, no.~3, pp.
  102--114, May 2010.

\bibitem{Shor_1985}
N.~Z. Shor, \emph{Minimization Methods for Non-Differentiable Functions}, ser.
  Springer Series in Computational Mathematics.\hskip 1em plus 0.5em minus
  0.4em\relax Springer-Verlag Berlin Heidelberg, 1985.

\bibitem{Bazaraa_2006}
M.~S. Bazaraa, H.~D. Sherali, and C.~M. Shetty, \emph{Nonlinear Programming:
  Theory and Algorithms}, 3rd~ed.\hskip 1em plus 0.5em minus 0.4em\relax
  Wiley-Interscience, 2006.

\bibitem{Bertsekas_1982}
D.~P. Bertsekas, ``Projected newton methods for optimization problems with
  simple constraints,'' \emph{{SIAM} J. Control Optim.}, vol.~20, no.~2, pp.
  221--246, Mar. 1982.

\bibitem{Kim_2010}
D.~Kim, S.~Sra, and I.~S. Dhillon, ``Tackling box-constrained optimization via
  a new projected quasi-newton approach,'' \emph{SIAM J. Sci. Comput.},
  vol.~36, no.~6, pp. 3548--3563, Dec. 2010.

\bibitem{Schmidt_2012}
M.~Schmidt, D.~Kim, and S.~Sra, ``Projected newton-type methods in machine
  learning,'' in \emph{Optimization for Machine Learning}, S.~Sra, S.~Nowozin,
  and S.~J. Wright, Eds.\hskip 1em plus 0.5em minus 0.4em\relax The {MIT}
  Press, 2012, ch.~11, pp. 305--329.

\bibitem{Richards_2005}
A.~A. Richards, \emph{Fundamentals of radar signal processing}.\hskip 1em plus
  0.5em minus 0.4em\relax New York, NY, {USA}: Mc{G}raw-Hill, 2005.

\bibitem{Horn_Johnson_1985}
R.~A. Horn. and C.~R. Johnson, \emph{Matrix Analysis}.\hskip 1em plus 0.5em
  minus 0.4em\relax Cambridge University Press, 1985.

\end{thebibliography}

\end{document}